\documentclass[12pt]{article}
\usepackage{xcolor}
\usepackage{amsmath}
\usepackage{amssymb}
\usepackage{amsfonts}
\usepackage{rotfloat}
\usepackage{graphicx}
\usepackage{mathrsfs}
\usepackage{array}
\usepackage{float}
\usepackage{textcomp}
\usepackage{multirow}
\usepackage{subfig}

\begin{document}
\renewcommand{\vec}{\pmb}
\newcommand{\dml}[1]{\bar{#1}}
\newcommand{\fdml}[1]{\tilde{#1}}
\definecolor{note}{rgb}{0.0, 0.53, 0.74}
\newcommand{\note}[1]{\textcolor{note}{#1}}
\renewcommand{\d}[1]{\mathrm{d}#1}
\newcommand{\bigO}[1]{O{\left(#1\right)}}
\newcommand{\citereq}{\note{[\todo{}]}}
\newcommand{\Rey}{\mbox{\textit{Re}}}
\newcommand{\Eo}{\mbox{\textit{Eo}}}
\newcommand{\Fr}{\mbox{\textit{Fr}}}
\newcommand{\Cn}{\mbox{\textit{Cn}}}
\newcommand{\Pe}{\mbox{\textit{Pe}}}
\newcommand{\Pra}{\mbox{\textit{Pr}}}
\newcommand{\We}{\mbox{\textit{We}}}
\newcommand{\unit}[1]{\hat{#1}}
\newcommand{\ie}{\textit{ie}.\@}
\newcommand{\eg}{\textit{eg}.\@}
\newcommand{\cf}{\textit{cf}.\@}
\newcommand{\todo}{\textsc{todo}}
\newcommand{\fdw}{\omega_{\text{dw}}}
\newcommand{\eq}{\text{eq}}
\newcommand{\avg}{\text{avg}}
\newcommand{\inv}{\text{inv}}
\newcommand{\init}{\text{init}}
\newcommand{\Ravg}{\langle R \rangle}
\newcommand{\hide}[1]{}

\title{Grand-potential phase field simulations of droplet growth and sedimentation 
in a two-phase ternary fluid}


\author{Werner Verdier$^{1,2}$, Alain Cartalade$^1$, Mathis Plapp$^2$\footnote{Corresponding author, mathis.plapp@polytechnique.fr}}
\date{\footnotesize $^1$ Université Paris-Saclay, CEA, Service de Thermo-hydraulique\\
et de Mécanique des Fluides, 91191 Gif-sur-Yvette, France\\
$^2$ Laboratoire de Physique de la Matière Condensée, CNRS,
École Polytechnique, \\Institut Polytechnique de Paris, 91120 Palaiseau, France}

\maketitle

\begin{abstract}
A methodology is built to model and simulate the dynamics of domain coarsening
of a two-phase ternary liquid with an arbitrary phase diagram. High numerical 
performance is obtained through the use of the phase field-method for interface
capturing, a lattice Boltzmann method numerical scheme for all the model equations, 
and a portable, parallel simulation code running on multiple GPUs. The model is 
benchmarked against an analytic solution for a ternary diffusion couple. 
It also reproduces the well-known power law for droplet coarsening during
Ostwald ripening without fluid flow. Large-scale simulations with flow illustrate 
the effects of momentum transport and buoyancy, as well as droplet coalescence
and sedimentation. 
\end{abstract}


\newpage

\section{Introduction}
When a homogeneous solid or liquid mixture is rapidly quenched into a thermodynamically 
unstable state, phase separation occurs. After an initial stage of spinodal decomposition,
during which microscopic fluctuations are amplified with time, spatial domains of the new 
equilibrium phases form and are separated by well-defined interfaces with a characteristic thickness. 
At later times, domain coarsening occurs: under the driving force of capillarity, the total interface
area is progressively reduced by the elimination of geometric features with high 
curvature, such as protrusions or small domains.

This phenomenon has been extensively studied because it provides an example of a system 
that never reaches equilibrium, but instead exhibits scaling laws \cite{gunton1983phase}. 
Indeed, in a system of infinite size, domain coarsening continues indefinitely 
and leads to patterns that are scale-invariant, that is, the structure and geometry 
of the pattern is statistically invariant in time, and only its overall scale grows 
with time as a power law. The growth exponent depends on the underlying transport 
processes (diffusion, hydrodynamics \cite{siggia_1979}, or elasticity)
and on the nature of the order parameter that describes the domain (scalar, vector or tensor,
conserved or non-conserved) \cite{Bray94,Bray02}.

Besides its theoretical interest, phase separation is also important for a large number
of industrially relevant processes, such as the formation of porous glasses and membranes.
One example that motivated the present work is the conditioning of nuclear waste in glass: the radioactive substances are mixed with a glass matrix, a mixture of glass formers 
that is optimized for long-term resistance to environmental stresses \cite{gin2017radionuclides}.
In order to increase the amount of waste, glass ceramics are being studied as an alternative
to conventional glass. For a simplified ternary system, a liquid-liquid phase separation can occur
when its global composition lies within a miscibility gap. The composition of the first liquid phase is representative of waste,
whereas the composition of the second one is representative of the glass matrix \cite{Schuller_etal_JACS2011,Pinet_etal_JNM2019}.

For the study of generic features of phase separation, such as spinodal decomposition
dynamics and scaling laws in coarsening, the Cahn-Hilliard equation \cite{Cahn58} and its 
various generalizations have been extensively used. Its appealing feature is that it can be
simply derived from out-of-equilibrium thermodynamics, taking as a starting point
the description of heterogeneous systems by a free-energy functional. The Cahn-Hilliard
equation can describe the complete time evolution of a phase-separating system, from
the initial spinodal decomposition to the late-stage coarsening, if transport is
governed by diffusion. It can be coupled with the Navier-Stokes 
equation \cite{brackbill_kothe_zemach_1992,jacqmin_1999} to model 
coarsening that is mediated by hydrodynamics, as well as the crossover between different 
coarsening regimes \cite{henry_tegze_2018,henry_tegze_2019}.

While the Cahn-Hilliard equation is thus an excellent tool to explore generic features
of phase-separating systems, it is rather difficult to adapt it for the description
of specific materials. There are two reasons for this. Firstly, the Cahn-Hilliard
equation is formulated for a binary system, with a concentration as the only dynamic
variable, and a free-energy functional that consists of a double-well potential and a
gradient term. In this formulation, the interface free energy is controlled by the interplay 
of these two terms. It is therefore difficult to modify the bulk thermodynamic properties, 
which are set by the curvature of the free-energy function, while leaving the interface energy 
invariant \cite{semprebon_ciro_kruger_2016}. Secondly, for multi-component
mixtures, there are several independent chemical compositions, and the gradient energy
coefficient becomes a symmetric matrix, with entries that are in principle determined
by the pairwise interaction between the different components. However, since those are
not directly known, the system is underdetermined, that is, for a target interface
energy, there are many possible choices for the gradient coefficients. This choice
has non-trivial consequences for the interface structure: interface adsorption of
certain components can occur, which modifies the interface energy and the macroscopic 
conservation laws for moving interfaces \cite{rasolofomanana_et_al_2022}.

An alternative way for describing interfaces in multicomponent mixtures has been
developed in the framework of phase-field theory \cite{ProvatasElder,Steinbach09,PlappHandbook}. 
The interface is described by a
scalar phase field, which can be seen as a smoothed indicator function (with 1
and 0 corresponding to presence or absence of the phase). The equation of motion
for this field derives from a Ginzburg-Landau free energy functional and is 
coupled to the concentration fields. The latter are then governed by their own 
free energy functions, which do not come into play for the determination of the 
interface properties, and which can therefore be chosen at will. The connection
between the phase-field model and the more traditional free-boundary problem can
be made using the well-established technique of matched asymptotic expansions
\cite{karma_rappel_1998,almgren_1999,echebarria_et_al_2004,badillo_2012}. 

For the quantitative description of moving interfaces in multi-component mixtures, 
the grand-potential model
(which is a reformulation of the earlier Kim-Kim-Suzuki model \cite{kim_kim_suzuki_1999}) 
has been widely used and benchmarked \cite{plapp_2011_2,Choudhury12,Plapp16}. 
It corresponds to a grand-canonical formulation of the
mixture thermodynamics, with the chemical potentials as basic variables, which 
is suitable because two phases that coexist in a mixture have different 
compositions but equal chemical potentials.

Here, we couple a grand-canonical phase-field model for multicomponent mixtures
with a description of fluid flow using the lattice-Boltzmann method 
(LBM) \cite{the_lattice_boltzmann_method} to achieve 
a model for domain coarsening that can be used for specific substances with given 
thermodynamics (obtained, for example, from a CALPHAD database \cite{calphad}). This model 
cannot describe the initial stages of spinodal decomposition, because the grand-potential 
formulation requires a monotonic relation between composition and chemical potentials, but it
provides a quantitative description of the late-stage domain coarsening regime. Since, in
real substances, the two phases generally do not have the same density, we include
buoyancy in the model, which makes it possible to describe sedimentation of droplets
of the heavier phase.

We develop an implementation in which all the equations, including the phase-field
model, are solved within the LBM formalism \cite{the_lattice_boltzmann_method}. 
Indeed, while LBM was initially developed for fluid flow, it was soon recognized 
that it is a general method for the time integration of partial differential 
equations, and we have previously used this formulation for solidification 
problems \cite{cartalade2016}. The advantage is that the simulation
code is entirely formulated in the LBM framework, and therefore high-performance
algorithms developed for LBM can be used. In particular, LBM is well adapted
for GPU parallelization.

In the present manuscript, we will first describe the grand-potential phase-field
model we use, which is similar to previously published work. Specifically, we
model a two-phase three-component fluid. We will then detail our
numerical implementation, and present a detailed benchmark study on a ternary 
diffusion couple, which shows that non-trivial interface conditions can be accurately
resolved by our formulation. Then, we present simulations of domain coarsening,
in two dimension without fluid flow as a benchmark, and in three dimensions
with fluid flow, in a large system which contains many droplets. These simulations
are carried out with the high-performance computing code 
LBM\_Saclay \cite{verdier_kestener_cartalade_2020}. These illustrative simulations 
demonstrate that our code can be used to perform large-scale studies, and thus is a 
suitable tool for the investigation of practical problems.

\section{Model}

\subsection{Thermodynamics}
We consider a ternary mixture, and denote by $c^A(\vec{x},t)$, $c^B(\vec{x},t)$,
and $c^C(\vec{x},t)$ the local molar fractions of the $A$, $B$, and $C$ components.
Since $c^A+c^B+c^C=1$, only two concentration fields are independent; we choose
to eliminate $c^C$ by writing $c^C=1-c^A-c^B$, and write the (Helmholtz) free energy
density of a homogeneous system as $f(c^A,c^B)$. For a specific substance, this
free energy function can be typically obtained from a CALPHAD \cite{calphad} database.
Such a database usually gives the molar Gibbs free energy, but we will suppose that
the molar volume is constant and independent of the composition, so that Gibbs and
Helmholtz free energies are equivalent. The variables that are thermodynamically
conjugate to the compositions are the diffusion potentials, that is, the difference 
of the chemical potentials of $A$ and $C$ and $B$ and $C$, respectively,
\begin{equation}
\mu^\alpha = V_a \frac{\partial f}{\partial c^\alpha}\qquad (\alpha=A,B)
\label{eq_mudef}
\end{equation}
where $V_a$ is the atomic volume (the molar volume divided by Avogadro's number). 
This factor has been included to give the diffusion potentials (which will also be 
called chemical potentials in the following) their usual dimension of energy \cite{Plapp16}.

For a phase-separating system, the ``free energy landscape'' $f(c^A,c^B)$ has a 
double-well structure, with two convex regions separated by a concave one; the latter
corresponds to the compositions for which a homogeneous system is thermodynamically
unstable. At equilibrium, the system is in a coexistence of two phases, called
``phase 0'' and ``phase 1'' in the following, which have different compositions,
each one being located in one of the wells of the free energy function.
Minimization of the free energy under the constraint of global mass conservation
yields the conditions for phase coexistence:
\begin{align}
	& V_a \left. \dfrac{\partial f_0}{\partial c^\alpha} \right|_{c^{\alpha,\eq}_0}
	= V_a \left. \dfrac{\partial f_1}{\partial c^\alpha} \right|_{c^{\alpha,\eq}_1} =
	\mu^{\alpha,\eq}, \\
	\begin{split}
		& f_0(c^{A,\eq}_0, c^{B,\eq}_0) - f_1(c^{A,\eq}_1,
		c^{B,\eq}_1) \\
		& \qquad = \dfrac{1}{V_a} \sum_{\alpha=A,B} \mu^{\alpha,\eq}
		(c^{\alpha,\eq}_0 - c^{\alpha,\eq}_1) ,
	\end{split}
	\label{eq:common_tangent_f}
\end{align}
where $\mu^{\alpha,\eq}$, $c_0^{\alpha,\eq}$, and $c_1^{\alpha,\eq}$ are the chemical
potentials and the compositions of phase 0 and 1 at two-phase equilibrium.
Geometrically, this corresponds to a ``common tangent plane'' to the free energy landscape.
From this geometric picture, or from a simple counting of degrees of freedom (three equations
for four unknown equilibrium compositions), it is clear that the solution to these
equations is not unique. In the space of compositions, which is usually visualized in
the Gibbs simplex, the various possible equilibria are indicated by {\em tie lines}
(see Figure~\ref{fig:diagram_ideal} below).
The global equilibrium corresponds to the unique tie line which contains the
composition inventory (the average composition) of the closed system.

The conditions for phase coexistence can be conveniently reformulated in the 
grand-canonical framework for open system, in which the fundamental variables
are the diffusion potentials, and the relevant thermodynamic potential is the
grand potential,
\begin{equation}
\omega (\mu^A, \mu^B) = f(c^A(\mu^A, \mu^B), c^B(\mu^A,\mu^B))
		 - \dfrac{1}{V_a} \sum_{\alpha=A,B} \mu^\alpha c^\alpha(\mu^A, \mu^B),
\label{eq:grandpotential}
\end{equation}
which is the Legendre transform of the free energy. In writing down this formula,
we have supposed that the functions $\mu^\alpha(c^A,c^B)$ can be inverted to yield
$c^\alpha(\mu^A,\mu^B)$. This is guaranteed only if these functions are monotonous,
which corresponds to a convex free energy landscape. In the case of a phase-separating
system, the Legendre transform must be taken separately for each convex region in
composition space (see below for details), which yields two different grand potential
functions for the two phases, $\omega_0$ and $\omega_1$. Phase coexistence is then 
simply given by
\begin{equation}
\omega_0(\mu^A,\mu^B)=\omega_1(\mu^A,\mu^B).
\end{equation}
This defines a coexistence line in the space of the intensive variables (the two diffusion
potentials), each point of which corresponds to a tie line in composition space.

In general, the calculation of the Legendre transform in Eq.~(\Ref{eq:grandpotential})
cannot be performed analytically, since free energy functions typically involve both polynomials and logarithms of the concentrations. It can always be performed numerically, but in 
the case of interest here
we can exploit the fact that during the late stages of the phase separation 
process, the concentrations will be close to the global equilibrium concentrations.
Therefore, only the vicinity in concentration space of the reference equilibrium
is relevant, and we can perform a second-order Taylor expansion of the free energy
around the equilibrium concentrations for each phase, 
as was already done in previous works \cite{Choudhury12},
\begin{eqnarray}
f_\pi(c^A,c^B)&=&f_\pi(c^{A,\eq}_\pi,c^{B,\eq}_\pi)+\sum_\alpha \frac{\mu^{\alpha,\eq}}{V_a}(c^\alpha-c^{\alpha,\eq}_\pi)\nonumber\\
&&\mbox{} + \frac{1}{2} \sum_{\alpha,\beta} K^{\alpha\beta}_\pi(c^\alpha-c^{\alpha,\eq}_\pi)(c^\beta-c^{\beta,\eq}_\pi).
\label{eq:Taylor}
\end{eqnarray}
Here and in the remainder of the text, exponents in Greek letters
and the index $\pi $ identify components $A$ or $B$, and phase $0$ or $1$, respectively.
In Eq.~(\ref{eq:Taylor}), we have used Eq.~(\Ref{eq_mudef}) in the first order terms, and 
the second order coefficients are given by
\begin{equation}
K^{\alpha\beta}_\pi=\left.\frac{\partial^2 f}{\partial c^\alpha \partial c^\beta}\right|_{c^{A,\eq}_\pi,c^{B,\eq}_\pi}.
\end{equation}
In this quadratic approximation, Eq.~(\ref{eq_mudef}) yields a linear relation between the
chemical potentials and the compositions,
\begin{equation}
\mu^\alpha_\pi(c^A,c^B)=\mu^{\alpha,\eq}+K^{\alpha A}_\pi(c^A-c^{A,\eq}_\pi)
                                         +K^{\alpha B}_\pi(c^B-c^{B,\eq}_\pi).
\end{equation}
This equation can easily be inverted to obtain $c^\alpha(\mu^A,\mu^B)$, which, together with Eq.~(\ref{eq:grandpotential}), yields an anaytic approximation for the grand potential.

The procedure outlined above is valid in the late stage of phase separation for arbitraty
free energy functions. Since, in the present contribution, we do not intend to model a
particular material, but are rather interested in benchmark calculations, we choose as
an example a simple model system, in which the matrix of second order coefficients is 
diagonal, with equal values for the two phases,
\begin{equation}
K_\pi^{\alpha\beta} = K \delta_{\alpha\beta},
\end{equation}
where $\delta_{\alpha\beta}$ is the Kronecker symbol. This corresponds to circular parabolic 
free energy wells with equal curvatures for both phases.
Furthermore, we will switch to the
new variables $\tilde\mu^\alpha=\mu^\alpha-\mu^{\alpha,\eq}$ and 
$\tilde\omega_\pi=\omega_\pi(\mu^A,\mu^B)-\omega_\pi(\mu^{A,\eq},\mu^{B,\eq})$, which amounts
to choosing the reference values for the chemical potentials and the grand potential.
In these variables, the grand potentials can be expressed as
\begin{equation}
	\tilde\omega_\pi ({\tilde\mu}^A, {\tilde\mu}^B)=- \dfrac{1}{2 K V_a^{2}} \sum_{\alpha=A,B} \left({{\tilde\mu}^\alpha}\right)^{2}- \dfrac{1}{V_a} \sum_{\alpha=A,B} {\tilde\mu}^\alpha c^{\alpha,\eq}_\pi.
\end{equation}
The phase diagram obtained from this model with $c^{A,\eq}_0=c^{B,\eq}_0=0.3$ and
$c^{A,\eq}_1=c^{B,\eq}_1=0.4$ is displayed in Figure~\ref{fig:diagram_ideal}.

\begin{figure}
	\centering
	\includegraphics[width=0.8\linewidth]{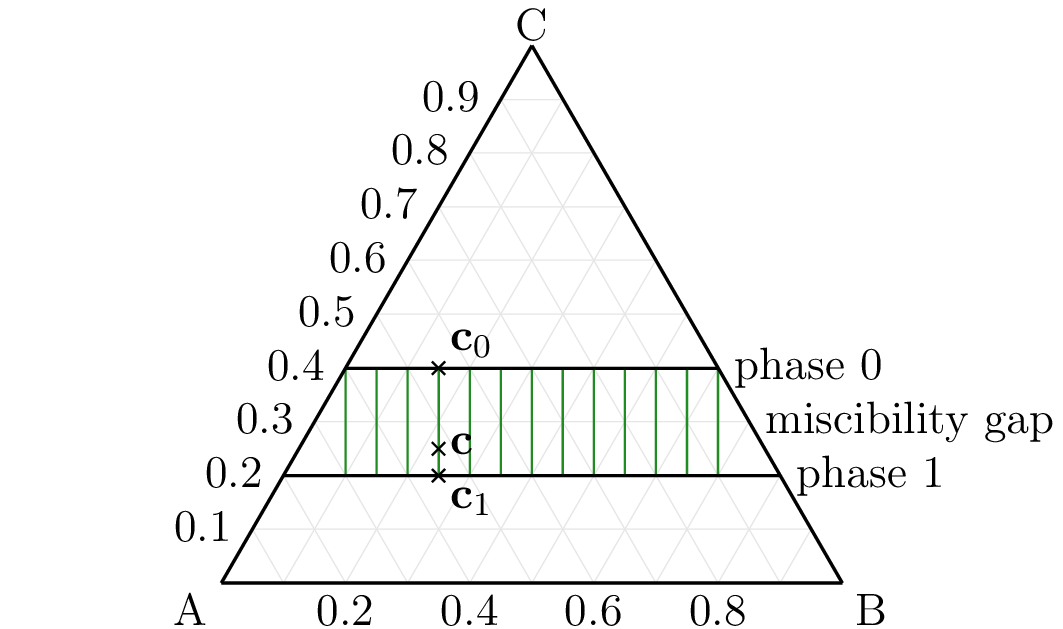}
	\caption{Ternary phase diagram of the idealized material considered in
	this work. It arises from the quadratic free energies with identical
	second derivatives for each phases. The horizontal black lines are the
	phase boundaries and the green lines are some of the tie-lines.
	Each of these tie-lines defines a valid chemical
	equilibrium with the compositions of each phase located at each end
	of the line.}
	\label{fig:diagram_ideal}
\end{figure}

\subsection{Grand potential phase field formalism}
In a multi-component Cahn-Hilliard model, the free energy landscape defined above
would be supplemented by square gradient terms in the free energy functional
to capture the free energy cost of inhomogeneities, which then adds non-local terms
(containing Laplace operators) to the diffusion potentials. Instead, we choose to
describe phases and interfaces by an additional scalar phase field
$\varphi(\vec{x}, t)$, with values in $[0, 1]$, whose extrema identify one of
the two bulk liquid phase. The smooth variations of $\varphi$ will locate the
diffuse interfaces.

The model is based on a phenomenological grand potential functional
\cite{plapp_2011_2,Choudhury12,Plapp16} over the volume $V$ of the system, 
function of the phase field $\varphi$ and of the fields of intensive thermodynamical 
variables $\mu^A$, $\mu^B$,
\begin{equation}
\Omega[\varphi, \mu^A, \mu^B] = 
   \int_V \left(\omega_{\rm int}(\varphi)+\omega_{\rm th}(\varphi,\mu^A,\mu^B)\right).
\label{eq:functional}
\end{equation}
The two contributions describe the interfacial and bulk (thermodynamic) contributions to
the total grand potential, respectively.

The interfacial part,
\begin{equation}
\omega_{\rm int}(\varphi) = \dfrac{\zeta}{2} |\nabla\varphi|^{2} + H\fdw(\varphi),
\end{equation}
contains a square gradient energy term and the double-well function $\fdw(\varphi) =
8\varphi^{2}(1-\varphi)^{2}$. They are parametrized by the
characteristic energy scales $H$ ($[E]  \cdot  L^{-3}$, height of the double-well)
and $\zeta$ ($[E]  \cdot  L^{-1}$, gradient energy coefficient). The system
evolves to minimize $\Omega$: the double-well term, favoring a sharp profile
for $\varphi$ at the interface, will then play against the gradient energy
term, which favors smooth variations of $\varphi$.
Assuming a plane interface at equilibrium (where the bulk term $\omega_{\rm th}$ is
absent), the interface solution is given by the hyperbolic tangent profile
\begin{equation}
	\varphi_0(x) = \dfrac{1}{2} \left( 1 + \tanh{\left(2x / W\right)} \right),
	\label{eq:tanh}
\end{equation}
with
\begin{equation}
	W = \sqrt{\zeta / H}
\end{equation}
the characteristic interface width. Reintroducing this profile in the integral
(\ref{eq:functional}) gives a characteristic value of the surface tension
due to the diffuse interface,
\begin{equation}
	\sigma = \dfrac{2}{3} HW .
\end{equation}
This relation also identifies $H$ as the characteristic energy-per-volume
scale of the excess free energy stored in the diffuse interface.
More details about these calculations can be found in introductory
texts about the phase-field method, for example in Refs.~\cite{LangerBegRohu,ProvatasElder,Plapp16}.

The bulk contribution is an interpolation between the grand potentials of the
two phases,
\begin{equation}
	\omega(\varphi, \mu^A, \mu^B) = \left[1-p(\varphi)\right] \omega_0(\mu^A, \mu^B) + p(\varphi) \omega_1(\mu^A,\mu^B) ,
\end{equation}
with $p(\varphi)=3\varphi^2-2\varphi^3$ an interpolation function odd around $\varphi = 1/2$, 
satisfying $p(\{0, 1\}) = \{0, 1\}$ and $p'(\{0, 1\}) = 0$, that can be seen as a smoothed
step function \cite{Plapp16}.

The interface-tracking equation, also known as the Allen-Cahn equation, is obtained
by relating linearly the time evolution of $\varphi$ and the decrease in $\Omega$,
expressed as the variational derivative,
\begin{equation}
	\partial_t \varphi = - M_\varphi \dfrac{\delta \Omega}{\delta \varphi}.
	\label{eq:allen_cahn}
\end{equation}
Here, $M_\varphi$ is a kinetic coefficient (phase-field mobility). The terms in this
equations that are generated from the interfacial energy stabilize the interface
profile, whereas the bulk term creates a driving force for interface motion if
the grand potentials of the two phases differ.

It is convenient to remove the energy dimensions: define the dimensionless 
chemical potentials $\dml{\mu}^\alpha = {\tilde\mu}^\alpha/ (K V_a)$, and
\begin{equation}
	\begin{split}
		\dml{\omega}_\pi (\dml{\mu}^A, \dml{\mu}^B) & = {\tilde\omega}_\pi ({\tilde\mu}^A,
		{\tilde\mu}^B)/K \\
		& = - \dfrac{1}{2} \sum_{\alpha=A,B} \left( ({\dml{\mu}^\alpha})^{2} +
		2 \dml{\mu}^\alpha c^{\alpha,\eq}_\pi \right).
	\end{split}
	\label{eq:dimless_grand_potentials}
\end{equation}
We can now write the interface tracking equation (\ref{eq:allen_cahn})
in terms of the fields $\varphi$, $\dml{\mu}^A$ and $\dml{\mu}^B$ as
\begin{equation}
	\tau_\varphi \partial_t \varphi = W^2 \nabla^{2}\varphi - \fdw'(\varphi) + \lambda p'(\varphi) \Delta \dml{\omega}(\dml{\mu}^A,
		\dml{\mu}^B)
\end{equation}
with the phase-field relaxation time $\tau_\varphi=1/(M_\varphi H)$ and the difference of the 
grand potential densities
\begin{equation}
	\begin{split}
		\Delta \dml{\omega} & = \dml{\omega}_0 - \dml{\omega}_1 \\
		& = - \sum_{\alpha=A,B} \dml{\mu}^\alpha \left( c^{\alpha,\eq}_0 -
		c^{\alpha,\eq}_1 \right) ,
	\end{split}
\end{equation}
where the thermodynamical coupling parameter is $\lambda = K / H$. We also define the scaled
phase-field mobility $\dml{M_\varphi}=M_\varphi H=W^2/\tau_\varphi$, which has the dimension
of a diffusion coefficient.

\subsection{Species diffusion and mixed formulation}

In the absence of hydrodynamic flow, the redistribution of chemical species 
occurs by diffusion. The concentration of each species is a locally conserved
quantity, hence
\begin{equation}
\partial_t c^\alpha(\vec{x},t) = - \nabla\cdot \vec{j}^\alpha(\vec{x},t).
\label{eq:conservation}
\end{equation}
According to the thermodynamics of irreversible processes, the species currents write
\begin{equation}
\vec{j}^\alpha(\vec{x},t)=-\sum_{\beta=A,B} M^{\alpha\beta}\nabla\mu^\beta,
\end{equation}
where $M^{\alpha\beta}$ are the components of the atomic mobility matrix.
Furthermore, in the grand-canonical setting, where the natural variables are
the chemical potentials, the composition can be obtained from the grand-potential
functional by
\begin{equation}
	c^\alpha = - V_a\dfrac{\delta  \Omega}{\delta  \mu^\alpha} = - V_a\left( p(1-\varphi)
	\dfrac{\partial\dml{\omega}_0}{\partial\dml{\mu}^\alpha} + p(\varphi)
	\dfrac{\partial\dml{\omega}_1}{\partial\dml{\mu}^\alpha} \right) .
	\label{eq:species_closure}
\end{equation}

In the original grand-potential method for a binary alloy \cite{plapp_2011_2}, the 
concentration was eliminated from Eq.~(\Ref{eq:conservation}) in favor of the chemical 
potential; however, for multi-component alloys, this requires numerous matrix 
inversions \cite{Choudhury12}. Therefore, we prefer
to use a mixed formulation, as in some previous works \cite{bayle_2020,bayle_2020_phd}: we keep
both the concentrations and the compositions as dynamic variables.
This allows to solve Eq.\@ (\ref{eq:conservation}), using the explicit algorithm of
one's choice to update $c^\alpha$ ; $\mu^\alpha$ is then updated by inverting
Eq.\@ (\ref{eq:species_closure}). Remark that the inversion step has a simple
analytical expression with the quadratic free energies, namely
\begin{equation}
	\dml{\mu}^\alpha = c^\alpha - p(1 - \varphi) c^{\alpha,\eq}_0 - p(\varphi) c^{\alpha,\eq}_1 .
\end{equation}
Here, we will use a diagonal mobility matrix
and remove the dimension of energy by defining $\dml{M}^\alpha=KM^{\alpha\alpha}$;
$\dml{M}^\alpha$ then has the dimension of a diffusion coefficient.
If the diffusion coefficient is different in the two phases, the mobility is
linearly interpolated, $\dml{M}^{\alpha}(\varphi)=\dml{M}_{0}^{\alpha}(1-\varphi)+\dml{M}_{1}^{\alpha}\varphi$ for both components $\alpha=A,B$.
Notice that in the bulk phases, Eq.\@ (\ref{eq:conservation}) reduces
to Fick's laws with fluxes $\dml{M}^\alpha \nabla c^\alpha$ because of the quadratic
free energies. This will be helpful in preliminary tests since analytical
solutions are available in this case (\cf{} sec. \ref{sec:diffusion_couple}).

\subsection{Equivalent sharp-interface model}
\label{sec:sharp_interface}

The fact that phase field models only track an implicit, diffuse interface
greatly facilitates the numerical treatment of two-phase problems. However,
the behavior of the physical variables other than $\varphi$ at the phase 
interface is at first glance unspecified. This is in contrast to the
sharp-interface formulations of free-boundary problems, whose well-posedness
depends on the presence of explicit boundary condition at the interface
for values and fluxes of the relevant transport fields.

An important effort in the phase field literature is made to bridge both
formulations by extracting implicit interface conditions from the phase field
equations in the asymptotic limit of small $W$ (“thin-interface limit"),
and to add corrections if necessary to obtain adjustable interface properties. 
This is done using the formalism of matched asymptotic analysis in terms of an asymptotic
parameter $\varepsilon$ expressing the scale separation between the interface thickness
and the physically relevant ``outer'' scales. As this process
is lengthy and has already been presented in much detail in other works,
we will only present its results here. The interested reader can refer to
\cite{almgren_1999} for a step-by-step detail of the calculations ; to
\cite{folch_et_al_1999} and \cite{badillo_2012} for a more detailed
definition of the curvilinear coordinates and subsequent reexpression of
the differential operators ; and to \cite{karma_rappel_1998} and then
\cite{echebarria_et_al_2004} for a discussion of what can be considered 
a “thin” interface.

The asymptotic analysis of the model without flow was realized
following the formalism of Almgren \cite{almgren_1999}. It was seen
that the case of a ternary system or the presence of a closure relation,
Eq.\@ (\ref{eq:species_closure}), introduce little novelty in the calculations.
The result is a Gibbs-Thomson relation for the grand potential at the
interface,
\begin{equation}
	\Delta \dml{\omega} = -\delta  \kappa - \beta V
	\label{eq:Gibbs-Thomson}
\end{equation}
with $\kappa$ the local curvature of the interface, $V$ its normal
velocity, and the associated coefficients are
\begin{align}
\delta & =\frac{2}{3}\frac{W}{\lambda}\label{eq:Delta}\\
\beta & =\frac{2}{3}\frac{W}{\lambda M_{\varphi}}-\frac{19}{120}\sum_{\alpha=A,B}\frac{W(c_{0}^{\alpha,\eq}-c_{1}^{\alpha,\eq})^{2}}{\dml{M}^{\alpha}}\label{eq:KineticCoeff}
\end{align}

The numerical factors are due to integrations across the diffuse interface
profile and depend on the choice of the double-well function and the interpolation
function $p(\varphi)$. The asymptotics also yields the composition balance at the interface,
\begin{equation}
V[c^{\alpha}]_{-}^{+}=-\left[\dml{M}^{\alpha}\partial_{n}\dml{\mu}^{\alpha}\right]_{-}^{+}\label{eq:MassBalance}
\end{equation}
where $[ \cdot ]^+_-$ denotes the jump of a field across the interface,
and $\partial_n$ the gradient projected on the interface normal. 

\subsection{Flow coupling}
To include hydrodynamic flow dynamics, we couple our phase-field model to the
incompressible Navier-Stokes equation,
\begin{align}
\boldsymbol{\nabla}\cdot\boldsymbol{u} & =0\label{eq:ConservMass}\\
\rho_{0}\left[\partial_{t}\boldsymbol{u}+\boldsymbol{\nabla}\cdot(\boldsymbol{u}\boldsymbol{u})\right] & =-\boldsymbol{\nabla}p_{h}+\boldsymbol{\nabla}\cdot\left[\rho_{0}\nu(\varphi)(\boldsymbol{\nabla}\boldsymbol{u}+\boldsymbol{\nabla}\boldsymbol{u}^{T})\right]+\boldsymbol{F}_{tot}\label{eq:QDM}\\
\partial_{t}\varphi+\boldsymbol{u}\cdot\boldsymbol{\nabla}\varphi & =\frac{W^2}{\tau_{\varphi}}\boldsymbol{\nabla}^{2}\varphi-\frac{1}{\tau_\varphi}\omega_{dw}^{\prime}(\varphi)+\frac{\lambda}{\tau_\varphi}p^{\prime}(\varphi)\Delta\dml{\omega}(\dml{\mu}^{A},\dml{\mu}^{B})\label{eq:PhaseField}\\
\partial_{t}c^{\alpha}+\boldsymbol{u}\cdot\boldsymbol{\nabla}c^{\alpha} & =\boldsymbol{\nabla}\cdot\left[\dml{M}^{\alpha}(\varphi)\boldsymbol{\nabla}\dml{\mu}^{\alpha}\right]\label{eq:Compos}\\
\dml{\mu}^{\alpha} & =c^{\alpha}-p(1-\varphi)c_{0}^{\alpha,\eq}-p(\varphi)c_{1}^{\alpha,\eq}\label{eq:PotChem}
\end{align}
where $\vec{u}$ is a phase-averaged velocity field \cite{sun_beckermann_2007,badillo_2012}, 
$\rho_{0}$ is the constant density, $p_{h}$ is the hydrodynamic
pressure enforcing the condition $\boldsymbol{\nabla}\cdot\boldsymbol{u}=0$
(Lagrange multiplier of that condition), $\nu(\varphi)$ is the (phase-dependent) kinematic
viscosity, the component index $\alpha$ is $A,B$ and the total force $\boldsymbol{F}_{tot}$ 
is the sum of the gravity force $\boldsymbol{F}_{g}=\varphi\Delta\rho\boldsymbol{g}$
in the Boussinesq approximation (with $\Delta\rho$ the density different between the
phases, phase 1 being denser than phase 0, and $\vec{g}$ the constant 
acceleration vector due to gravity) and the surface tension 
force $\boldsymbol{F}_{\sigma}$ defined by \cite{brackbill_kothe_zemach_1992,jacqmin_1999}:

\begin{equation}
\boldsymbol{F}_{\sigma}=\frac{3}{2}\sigma W\left[\frac{\omega_{dw}^{\prime}(\varphi)}{W^{2}}-\boldsymbol{\nabla}^{2}\varphi\right]\boldsymbol{\nabla}\varphi\label{eq:SurfaceTension_Force}
\end{equation}
For the equilibrium solution $\varphi=\varphi_{0}$, the term inside
the bracket is equivalent to $\kappa\bigl|\boldsymbol{\nabla}\varphi\bigr|$
where $\kappa$ is the curvature defined by $\kappa=-\boldsymbol{\nabla}\cdot\boldsymbol{n}$.
The surface tension force can be expressed by $\boldsymbol{F}_{s}=\delta_{d}\sigma\kappa\boldsymbol{n}$,
i.e. the quantity $\sigma\kappa\boldsymbol{n}$ is spread over $\delta_{d}=(3/2)W\bigl|\boldsymbol{\nabla}\varphi\bigr|^{2}$
where $W$ is the thickness of the diffuse interface. 

In the phase-field equation Eq. (\ref{eq:PhaseField}), for simplifying
the notations in the next Section, we define the following source
term:

\begin{equation}
\mathscr{S}_{\varphi}(\varphi,\dml{\mu}^{A},\dml{\mu}^{B})=-\frac{1}{\tau_{\varphi}}\omega_{dw}^{\prime}(\varphi)+\frac{\lambda}{\tau_{\varphi}}p^{\prime}(\varphi)\Delta\dml{\omega}(\dml{\mu}^{A},\dml{\mu}^{B})\label{eq:SourceTerm_Phi}
\end{equation}
which contains the contributions of the double-well (first term) and
the thermodynamic imbalance $\Delta\dml{\omega}$ (second term)
responsible for the displacement of the interface. For $\boldsymbol{u}=\boldsymbol{0}$,
the interface is displaced by diffusion until the thermodynamic equilibrium is
reached, i.e. when the grand-potential densities of each phase are
equal $\dml{\omega}_{0}(\dml{\mu}^{A},\dml{\mu}^{B})=\dml{\omega}_{1}(\dml{\mu}^{A},\dml{\mu}^{B})$.

\section{Lattice Boltzmann Method}
The numerical resolution of the model is done through
the LBM, which consists of a discretization of the Boltzmann equation
in phase-space. This defines a regular lattice, and the time-explicit
resolution of the equation is assimilable to the distribution functions
undergoing a collision step on the nodes followed by a transport step
along the edges. According to the choice of collision operator (here
the simple BGK operator), of equilibrium functions, and of additional
source and force terms, the moments of the distribution function thus
solved can be shown to be solution of conservative time-evolution
PDEs such as the ones of the present model. This is coherent with
a discretized Chapman-Enskog expansion.

LBM is a very popular method to simulate various conservative PDEs and more particularly the Navier-Stokes equations. For those latter ones, more traditional methods use a prediction-correction algorithm that requires solving a time-consuming Poisson equation. The LBM benefits from the efficiency of the equivalent articifial compressibility algorithm. The collision operator is local and each discrete distribution function follows an identical evolution equation. The method is therefore very simple to implement and is parallel in nature by efficiently exploiting the shared memory. Moreover, coupled with a distributed memory parallelism (e.g. MPI), that algorithm is very efficient. The interested reader can refer to references such as \cite{the_lattice_boltzmann_method} for more details on the LBM.

In this work, the standard D2Q9 and D3Q19 lattices are used for all 2D or 3D simulations,
respectively. Those lattices correspond to a discretization of the
velocity-space with 9 or 19 discrete velocities $\boldsymbol{c}_{k}$
defined by $\boldsymbol{c}_{k}=(\delta x/\delta t)\boldsymbol{e}_{k}$
with $\delta x$ the space step, $\delta t$ the time step, and $\boldsymbol{e}_{k}$
the direction vectors. Each set of directions is weighted by a scalar
value $w_{k}$. The directions $\boldsymbol{e}_{k}$ and their weights
are listed in Table \ref{tab:D2Q9} for D2Q9 and Table \ref{tab:D3Q19}
for D3Q19.

We also define a characteristic speed $c_{s}$ defined by $c_{s}=(1/\sqrt{3})\delta x/\delta t$.
The distribution functions and the discrete Lattice
Boltzmann equations (LBE) with the BGK collision operator are detailed
next in Section \ref{subsec:LBM_Flow} for fluid flow, in section
\ref{subsec:LBM_Phi} for phase-field and in Section \ref{subsec:LBM_Compos}
for the composition equations. Many alternative collision operators exist in LBM literature. The most popular operators are the "Two-Relxation-Times" (TRT) and "Multiple-Relaxation-Times" (MRT) which define respectively two (for TRT) or more (for MRT) additional collision rates to tune for improving stability and accuracy. Here, we present a proof of concept of a thermodynamic model that is based on the grand-potential functional and coupled with fluid flow. The ratio of diffusion coefficients are almost 1 and the kinematic viscosity of each phase is identical. The BGK operator is sufficient in this work.

\begin{table}
\begin{centering}
\subfloat[\label{tab:D2Q9}Lattice D2Q9.]{
\centering{}%
\begin{tabular}{lccccl}
\hline 
$k$ & \multicolumn{4}{c}{\textbf{$\boldsymbol{e}_{k}$}} & $w_{k}$\tabularnewline
\hline 
\hline 
$0$ & \multicolumn{4}{c}{$(0,0)$} & $4/9$\tabularnewline
$1,2,3,4$ & $(+1,0)$ & $(0,+1)$ & $(-1,0)$ & $(0,-1)$ & $1/9$\tabularnewline
$5,6,7,8$ & $(+1,+1)$ & $(-1,+1)$ & $(-1,-1)$ & $(+1,-1)$ & $1/36$\tabularnewline
\hline 
 &  &  &  &  & \tabularnewline
 &  &  &  &  & \tabularnewline
 &  &  &  &  & \tabularnewline
 &  &  &  &  & \tabularnewline
\end{tabular}}\hspace{1cm}\subfloat[\label{tab:D3Q19}Lattice D3Q19]{
\centering{}%
\begin{tabular}{lcccl}
\hline 
$k$ & \multicolumn{3}{c}{\textbf{$\boldsymbol{e}_{k}$}} & $w_{k}$\tabularnewline
\hline 
\hline 
$0$ & \multicolumn{3}{c}{$(0,0,0)$} & $1/3$\tabularnewline
\hline 
$1,2,3$ & $(+1,0,0)$ & $(-1,0,0)$ & $(0,+1,0)$ & \multirow{2}{*}{$1/18$}\tabularnewline
$4,5,6$ & $(0,-1,0)$ & $(0,0,+1)$ & $(0,0,-1)$ & \tabularnewline
\hline 
$7,8,9$ & $(+1,+1,0)$ & $(-1,-1,0)$ & $(+1,0,+1)$ & \multirow{4}{*}{$1/36$}\tabularnewline
$10,11,12$ & $(-1,0,-1)$ & $(0,+1,+1)$ & $(0,-1,-1)$ & \tabularnewline
$13,14,15$ & $(+1,-1,0)$ & $(-1,+1,0)$ & $(+1,0,-1)$ & \tabularnewline
$16,17,18$ & $(-1,0,+1)$ & $(0,+1,-1)$ & $(0,-1,+1)$ & \tabularnewline
\hline 
\end{tabular}}
\par\end{centering}
\caption{Detail of the direction vectors $\boldsymbol{e}_{k}$ and scalar weights
$w_{k}$ associated to each velocity of the D2Q9 (\ref{tab:D2Q9})
and D3Q19 (\ref{tab:D3Q19}) lattices.}
\end{table}

\subsection{\label{subsec:LBM_Flow}LBM for fluid flow}

For simulating the fluid flow, the numerical scheme works on the distribution
function $f_{k}(\boldsymbol{x},t)$ for which its evolution is given
by the lattice Boltzmann equation:

\begin{equation}
f_{k}(\boldsymbol{x}+\boldsymbol{c}_{k}\delta t,t+\delta t)=f_{k}(\boldsymbol{x},t)-\frac{1}{\tau_{f}(\varphi)+0.5}\left[f_{k}(\boldsymbol{x},t)-f_{k}^{eq}(\boldsymbol{x},t)\right]+\delta t\mathcal{F}_{k}(\boldsymbol{x},t)\label{eq:LBE_FluidFlow}
\end{equation}
where the relaxation rate $\tau_{f}$ is related to the kinematic
viscosity by $\nu(\varphi)=\tau_{f}(\varphi)c_{s}^{2}\delta t$ and
the source term $\mathcal{F}_{k}(\boldsymbol{x},t)$ contains the
total force term $\boldsymbol{F}_{tot}$ of Eq. (\ref{eq:QDM}). Several
forcing schemes exist in the LBM literature, here we choose one of
the two most popular methods \cite{He-Shan-Doolen_PRE-Rapid1998}:

\begin{align}
\mathcal{F}_{k}(\boldsymbol{x},t) & =\Gamma_{k}(\boldsymbol{x},t)(\boldsymbol{c}_{k}-\boldsymbol{u})\cdot\boldsymbol{F}_{tot}\label{eq:ForcingTerm}\\
\Gamma_{k}(\boldsymbol{x},t) & =w_{k}\left[1+\frac{\boldsymbol{c}_{k}\cdot\boldsymbol{u}}{c_{s}^{2}}+\frac{(\boldsymbol{c}_{k}\cdot\boldsymbol{u})^{2}}{2c_{s}^{4}}-\frac{\boldsymbol{u}^{2}}{2c_{s}^{2}}\right]\label{eq:GammaFunction}
\end{align}
The moment of order zero of Eq. (\ref{eq:ForcingTerm}) is $\sum_{k}\mathcal{F}_{k}=0$
and its moment of first-order is equal to $\sum_{k}\mathcal{F}_{k}\boldsymbol{c}_{k}=\boldsymbol{F}_{tot}$.
For recovering the incompressible Navier-Stokes equation with the
Chapman-Enskog expansion, the equilibrium distribution function $f_{k}^{eq}$
has to be defined by \cite{He-Luo_Incompressible_JSP1997}:

\begin{equation}
f_{k}^{eq}(\boldsymbol{x},t)=w_{k}\left[p_{h}+\rho_{0}c_{s}^{2}\left(\frac{\boldsymbol{c}_{k}\cdot\boldsymbol{u}}{c_{s}^{2}}+\frac{(\boldsymbol{c}_{k}\cdot\boldsymbol{u})^{2}}{2c_{s}^{4}}-\frac{\boldsymbol{u}^{2}}{2c_{s}^{2}}\right)\right]-\frac{\delta t}{2}\mathcal{F}_{k}\label{eq:Feq}
\end{equation}
where the term $\mathcal{F}_{k}\delta t/2$ has been substracted for
capturing the second-order accuracy when the external force term is
taking into account in Eq. (\ref{eq:LBE_FluidFlow}). That trick is
equivalent to a change of variables in the distribution function to conserve
an explicit algorithm after the trapezoidal integration of the Boltzmann
equation. Because of that change of variables, 
the term $\sum_{k}\mathcal{F}_{k}\boldsymbol{c}_{k}\delta t/2$
must be added for updating the velocity for each time-step. After
collision and streaming, the hydrodynamic pressure is obtained by
the moment of order zero of $f_{k}$, and the velocity by the moment
of first order:

\begin{align}
p_{h} & =\sum_{k}f_{k}\label{eq:HydroPressure}\\
\rho_{0}\boldsymbol{u} & =\frac{1}{c_{s}^{2}}\sum_{k}f_{k}\boldsymbol{c}_{k}+\frac{\delta t}{2}\boldsymbol{F}_{tot}\label{eq:Velocity}
\end{align}

When a high contrast of density exists between both phases, another
popular method is widely applied for defining the equilibrium distribution
function $f_{k}^{eq}$. In that case, the density becomes an interpolation
of the bulk density of each phase (e.g. $\varrho(\varphi)$), and
a dimensionless pressure $p^{\star}=p_{h}/\varrho(\varphi)c_{s}^{2}$
is introduced in the equilibrium Eq. (\ref{eq:Feq}) \cite{zu_he_2013}. In that
case, two supplementary forces must be added in $\boldsymbol{F}_{tot}$,
the pressure and viscosity forces \cite{Fakhari_etal_PRE2017}, in
order to recover the incompressible Navier-Stokes. Here, we assume
that the density is identical in both phases, so that the classical
equilibrium Eq. (\ref{eq:Feq}) is sufficient for the simulations.

The surface tension force Eq. (\ref{eq:SurfaceTension_Force}) requires
computing a gradient term and a laplacian term. The gradient is evaluated
by the directional derivatives:

\begin{equation}
\boldsymbol{e}_{k}\cdot\boldsymbol{\nabla}\varphi\bigr|_{\boldsymbol{x}}=\frac{1}{2\delta x}\left[\varphi(\boldsymbol{x}+\boldsymbol{e}_{k}\delta x)-\varphi(\boldsymbol{x}-\boldsymbol{e}_{k}\delta x)\right]\label{eq:DerivDirec_Grad}
\end{equation}
where the number of directional derivatives is equal to the number
of moving directions $\boldsymbol{e}_{k}$ on the lattice i.e. $N_{pop}$.
The gradient is obtained by:

\begin{equation}
\boldsymbol{\nabla}\varphi\bigr|_{\boldsymbol{x}}=3\sum_{k=0}^{N_{pop}}w_{k}\boldsymbol{e}_{k}\left(\boldsymbol{e}_{k}\cdot\boldsymbol{\nabla}\varphi\bigr|_{\boldsymbol{x}}\right).\label{eq:Grad}
\end{equation}

For the calculation of the $\boldsymbol{\nabla}^{2}\varphi$, all
directions of propagation are taken into account by

\begin{equation}
(\boldsymbol{e}_{k}\cdot\boldsymbol{\nabla})^{2}\varphi\bigr|_{\mathbf{x}}=\frac{1}{\delta x^{2}}\left[\varphi(\boldsymbol{x}+\boldsymbol{e}_{k}\delta x)-2\varphi(\boldsymbol{x})+\varphi(\boldsymbol{x}-\boldsymbol{e}_{k}\delta x)\right]\label{eq:DerivDirect_Laplacian}
\end{equation}
which are used to compute the laplacian:

\begin{equation}
\boldsymbol{\nabla}^{2}\varphi\bigr|_{\boldsymbol{x}}=3\sum_{k\neq0}w_{k}(\boldsymbol{e}_{k}\cdot\boldsymbol{\nabla})^{2}\varphi\bigr|_{\boldsymbol{x}}.\label{eq:Laplacian}
\end{equation}

\subsection{\label{subsec:LBM_Phi}LBM for phase-field}

For the phase-field equation, we introduce a new distribution function
$g_{k}(\boldsymbol{x},t)$ evolving with the LBE: 

\begin{equation}
g_{k}(\boldsymbol{x}+\boldsymbol{c}_{k}\delta t,t+\delta t)=g_{k}(\boldsymbol{x},t)-\frac{1}{\tau_{g}+0.5}\left[g_{k}(\boldsymbol{x},t)-g_{k}^{eq}(\boldsymbol{x},t)\right]+\delta t\mathcal{G}_{k}(\boldsymbol{x},t)\label{eq:LBE_Phi}
\end{equation}
Eq. (\ref{eq:PhaseField}) is an Advection-Diffusion type Equation
(ADE). The equilibrium distribution function $g_{k}^{eq}$ is designed
such as its moments of order zero, one, and two are respectively equal
to $\varphi$, $\boldsymbol{u}\varphi$ and $\overline{\overline{\boldsymbol{I}}}\varphi$
where $\overline{\overline{\boldsymbol{I}}}$ is the identity tensor
of second order:

\begin{equation}
g_{k}^{eq}(\boldsymbol{x},t)=w_{k}\varphi\left[1+\frac{\boldsymbol{c}_{k}\cdot\boldsymbol{u}}{c_{s}^{2}}\right]-\frac{\delta t}{2}\mathcal{G}_{k}\label{eq:Geq}
\end{equation}
The first term in the right-hand side of Eq. (\ref{eq:Geq}) is the
classical term to recover the ADE after the Chapman-Enskog procedure,
and the source term $\mathcal{G}_{k}$ is simply defined such as its
moment of order zero is equal to $\mathscr{S}_{\varphi}(\varphi,\dml{\mu}^{A},\dml{\mu}^{B})$:

\begin{equation}
\mathcal{G}_{k}=w_{k}\mathscr{S}_{\varphi}(\varphi,\dml{\mu}^{A},\dml{\mu}^{B})\label{eq:Source_LBE_Phi}
\end{equation}
where $\mathscr{S}_{\varphi}$, defined by Eq. (\ref{eq:SourceTerm_Phi}).
The interface mobility $\dml{M}_{\varphi}=W^2/\tau_{\varphi}$ is a constant which is related
to the relaxation rate $\tau_{g}$ by $\dml{M}_{\varphi}=\tau_{g}c_{s}^{2}\delta t$.
Finally, after collision and streaming, the phase-field $\varphi$
is updated at each time-step by:

\begin{equation}
\varphi=\sum_{k}g_{k}+\frac{\delta t}{2}\mathscr{S}_{\varphi}\label{eq:Moment0_Phi}
\end{equation}

\subsection{\label{subsec:LBM_Compos}LBM for composition equations}

For the treatment of the composition equations, we introduce two distribution functions
$h_{k}^{\alpha}(\boldsymbol{x},t)$ (for $\alpha=A,B$) which evolve
with the LBE:

\begin{equation}
h_{k}^{\alpha}(\boldsymbol{x}+\boldsymbol{c}_{k}\delta t,t+\delta t)=h_{k}^{\alpha}(\boldsymbol{x},t)-\frac{1}{\tau_{h}^{\alpha}(\varphi)+0.5}\left[h_{k}^{\alpha}(\boldsymbol{x},t)-h_{k}^{eq,\alpha}(\boldsymbol{x},t)\right]\label{eq:LBE_c_alpha}
\end{equation}
Eqs (\ref{eq:Compos}) look like advection-diffusion equations,
but the equilibrium functions $h_{k}^{eq,\alpha}$ must be defined
such that its moment of order two is equal to $\overline{\overline{\boldsymbol{I}}}\dml{\mu}^{\alpha}$
where $\dml{\mu}^{\alpha}$ is related to the composition $c^{\alpha}$
(moment of order zero) by Eq. (\ref{eq:PotChem}). In that relation,
$c_{0}^{\alpha,\eq}$ and $c_{1}^{\alpha,\eq}$ are two scalar input values which
are the two thermodynamic equilibrium compositions of each phase 0
and 1. Because the moments of order 2 differs from the moment of order
0, the equilibrium distribution function must be slightly modified
compared to the classical equilibrium for an ADE by:

\begin{equation}
h_{k}^{eq,\alpha}(\boldsymbol{x},t)=\begin{cases}
c^{\alpha}-(1-w_{0})\dml{\mu}^{\alpha} & \text{for }k=0\\
w_{k}\dml{\mu}^{\alpha}+w_{k}c^{\alpha}\frac{\boldsymbol{c}_{k}\cdot\boldsymbol{u}}{c_{s}^{2}} & \text{for }k\neq0
\end{cases}\label{eq:Heq}
\end{equation}
In that equilibrium distribution, the moment of order zero is given
by the first term of the first line and the moment of order one is
given by the second term of the second line. That method is extensively
used for simulating the advective Cahn-Hilliard equation (\cite{fakhari_rahimian_2010}).
The mobility coefficients $\dml{M}^{\alpha}(\varphi)$ are related to the relaxation rate $\tau_{h}^{\alpha}(\varphi)$ by $\dml{M}^{\alpha}(\varphi)=\tau_{h}^{\alpha}(\varphi)c_{s}^{2}\delta t$.
After collision and streaming, the composition is obtained by the
moment of order zero of $h_{k}^{\alpha}$:

\begin{equation}
c^{\alpha}=\sum_{k}h_{k}^{\alpha}\label{eq:Moment0_c_alpha}
\end{equation}

\section{Numerical simulations}\label{Simuls}

The methods of the previous section have been implemented in a C++ code: LBM\_Saclay \cite{LBMsaclaycode}. Its main feature is its multi-architecture portability by simple modifications
of compilation options in the makefile. This is possible thanks to the Kokkos library \cite{kokkos}. Thus the same code can be run either on multi-CPU partition or on multi-GPU partition of a supercomputer. Time performances of LBM\_Saclay have been compared between graphic cards and standard CPU in our previous paper \cite{verdier_kestener_cartalade_2020}. Because of the efficiency of the former ones, the simulations of this work have run on the multi-GPU partitions of the supercomputers Jean-Zay (IDRIS, France) and Topaze (CCRT, France).

\subsection{Ternary diffusion couple}
\label{sec:diffusion_couple}
We start by presenting a quantitative validation of the model and its
implicit Gibbs-Thomson condition for a plane interface without
flow. This demonstrates that the phase field model is able
to reproduce the known solution of a sharp-interface problem, the
symmetrical ternary diffusion couple. Phase field studies of this
problem already exist in the literature, and we use this as a test for our model \cite{heulens_blanpain_moelans_2011,lahiri_abinandanan_choudhury_2017}. An interface
splits the infinite one-dimensional space in two domains at $x_{I}(t)$, with
phase 0 on its left and phase 1 on its right. The interface is displaced
at velocity $\dot{x}_{I}(t)$ by the interdiffusion of the components. This 
problem can be stated as the one-dimensional free boundary problem

\begin{align}
\partial_{t}c^{\alpha} & =\dml{M}^{\alpha}\partial_{xx}^{2}\dml{\mu}^{\alpha}, & x<x_{I}(t)\text{ or }x>x_{I}(t)\label{eq:Stefan}\\
\dml{\mu}^{\alpha} & =\dml{\mu}_{int}^{\alpha}, & x=x_{I}(t)\label{eq:StefanInt1}\\
\dot{x}_{I}(c^{\alpha}\bigr|_{x_{I}^{-}}-c^{\alpha}\bigr|_{x_{I}^{+}}) & =-\dml{M}^{\alpha}(\partial_{x}c^{\alpha}\bigr|_{x_{I}^{-}}-\partial_{x}c^{\alpha}\bigr|_{x_{I}^{+}}) & x=x_{I}(t)\label{eq:StefanInt2}\\
\dml{\mu}^{\alpha} & =\dml{\mu}_{-\infty}^{\alpha} & x=-\infty\label{eq:Stefan-inf}\\
\dml{\mu}^{\alpha} & =\dml{\mu}_{+\infty}^{\alpha} & x=+\infty\label{eq:Stefan+inf}
\end{align}
with the initial conditions

\begin{equation}
x_{I}(0)=0; \qquad
\dml{\mu}^{\alpha}=\dml{\mu}_{-\infty}^{\alpha} \quad (x<0);\qquad
\dml{\mu}^{\alpha}=\dml{\mu}_{+\infty}^{\alpha} \quad (x>0).
\label{eq:CondInit}
\end{equation}

The problem corresponds to evaluating our phase field model in 1D
with $\varphi=0$ or $\varphi=1$ and taking the interface conditions
(\ref{eq:Gibbs-Thomson}), (\ref{eq:MassBalance}) with $\beta=0$ and $\kappa=0$. 
In the sharp-interface view, the relation between $\dml{\mu}^{\alpha}$ and $c^{\alpha}$
is linear, allowing the problem to be rewritten into the classical
Stefan problem in terms of either $c^{\alpha}$ or $\dml{\mu}^{\alpha}$
only. It is then known \cite{maugis_et_al_1997} that a self-similar
solution exists with

\begin{equation}
x_{I}(t)=\xi\sqrt{t}\label{eq:InterfPosition}
\end{equation}
and

\begin{equation}
\dml{\mu}^{\alpha}(x,t)=\begin{cases}
\dml{\mu}_{-\infty}^{\alpha}+(\dml{\mu}_{int}^{\alpha}-\dml{\mu}_{-\infty}^{\alpha})\frac{\text{erfc}(-x/2\sqrt{\dml{M}^{\alpha}t})}{\text{erfc}(-\xi/2\sqrt{\dml{M}^{\alpha}})} & -\infty<x<x_{I}(t)\\
\dml{\mu}_{+\infty}^{\alpha}+(\dml{\mu}_{int}^{\alpha}-\dml{\mu}_{+\infty}^{\alpha})\frac{\text{erfc}(x/2\sqrt{\dml{M}^{\alpha}t})}{\text{erfc}(\xi/2\sqrt{\dml{M}^{\alpha}})} & x_{I}(t)<x<+\infty
\end{cases}\label{eq:Solution_Mu}
\end{equation}

The statement and solution of the problem in terms of the
composition fields is analogous but with a discontinuity at the interface,
in coherence with chemical equilibrium. The couple of interface values
$(\dml{\mu}_{int}^{A},\dml{\mu}_{int}^{B})$ are a solution
of the system, meaning

\begin{equation}
\dml{\omega}_{0}(\dml{\mu}_{int}^{A},\dml{\mu}_{int}^{B})-\dml{\omega}_{1}(\dml{\mu}_{int}^{A},\dml{\mu}_{int}^{B})=0.
\end{equation}

A ternary system has access to a continuous set of such equilibrium
couples (phase diagram tie-lines). The additional dynamical constraints
select one particular equilibrium in this continuum: the coefficient $\xi$ is determined
as the solution of a transcendental equation that couples the thermodynamic
equilibrium and dynamical parameters, see Eq. (22) in~\cite{maugis_et_al_1997}.

\begin{sidewaystable}
\begin{centering}
\begin{tabular}{lllclllclll}
\cline{1-3} \cline{2-3} \cline{3-3} \cline{5-7} \cline{6-7} \cline{7-7} \cline{9-11} \cline{10-11} 
\multicolumn{3}{c}{\textbf{Domain}} & \hspace{5mm} & \multicolumn{3}{c}{\textbf{Phase-field parameters}} & \hspace{5mm} & \multicolumn{3}{c}{\textbf{Transport parameters}}\tabularnewline
\cline{1-3} \cline{2-3} \cline{3-3} \cline{5-7} \cline{6-7} \cline{7-7} \cline{9-11} \cline{10-11} \cline{11-11} 
Symbol & Value & Dim &  & Symbol & Value & Dim &  & Symbol & Value & Dim\tabularnewline
$[-L_{x},L_{x}]$ & $[-1,1]$ & {[}L{]} &  & $W$ & $1.2\times10^{-3}$ & {[}L{]} &  & $(c_{0}^{A,\eq},c_{0}^{B,\eq})$ & $(0.3,0.3)$ & {[}--{]}\tabularnewline
$[-L_{y},L_{y}]$ & $[-0.002,0.002]$ & {[}L{]} &  & $\lambda$ & $155.94541910\cong155.95$ & {[}--{]} &  & $(c_{1}^{A,\eq},c_{1}^{B,\eq})$ & $(0.4,0.4)$ & {[}--{]}\tabularnewline
$N_{x}$ nodes & $3000$ & {[}--{]} &  & $\dml{M}_{\varphi}$ & $1.2$ & {[}L{]}$^{2}$/{[}T{]} &  & $(\dml{M}^{A},\dml{M}^{B})$ & $(1,0.8)$ & {[}L{]}$^{2}$/{[}T{]}\tabularnewline
$N_{y}$ nodes & $6$ & {[}--{]} &  &  &  &  &  & $(c^{A},c^{B})_{-\infty}$ & $(0.4,0.175)$ & {[}--{]}\tabularnewline
$\delta x$ & $1/1500$ & {[}L{]} &  &  &  &  &  & $(c^{A},c^{B})_{+\infty}$ & $(0.225,0.6)$ & {[}--{]}\tabularnewline
$\delta t$ & $1.481481\times10^{-9}$ & {[}T{]} &  &  &  &  &  &  &  & \tabularnewline
\hline 
\end{tabular}
\par\end{centering}
\caption{\label{tab:Parameters_Diffusion-Couple}Parameters for the simulation
of symmetrical diffusion couple. $\lambda^{\star}$ is the value that
cancels the kinetic coefficient $\beta$ (Eq. (\ref{eq:KineticCoeff})).
In all simulations, that coefficient will be simply noted $\lambda$
and approximated by $155.95$. The values of $(c^{A},c^{B})_{-\infty}$
and $(c^{A},c^{B})_{+\infty}$ are respectively equivalent to $(\dml{\mu}^{A},\dml{\mu}^{B})_{-\infty}=(0.1,-0.125)$
and $(\dml{\mu}^{A},\dml{\mu}^{B})_{+\infty}=(-0.175,0.2)$.
The values of collision rates corresponding respectively to $\dml{M}_{\varphi}$, $\dml{M}^{A}$ and $\dml{M}^{B}$ are: $\tau_{g}=0.6333$, $\tau_{h}^{A}=0.6111$ and $\tau_{h}^{B}=0.5888$.}
\end{sidewaystable}

\begin{figure}
\begin{centering}
\includegraphics[height=5cm]{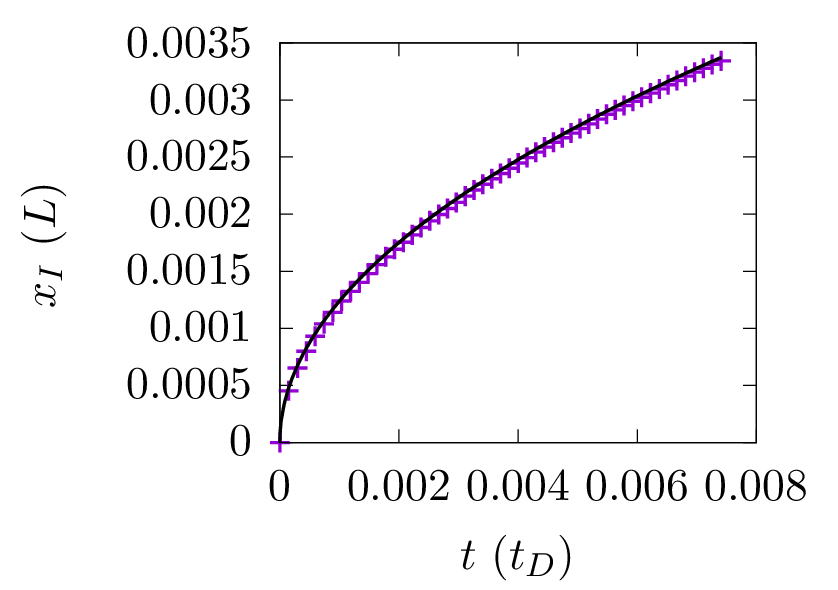}
\hspace{4.5mm}
\includegraphics[height=5cm]{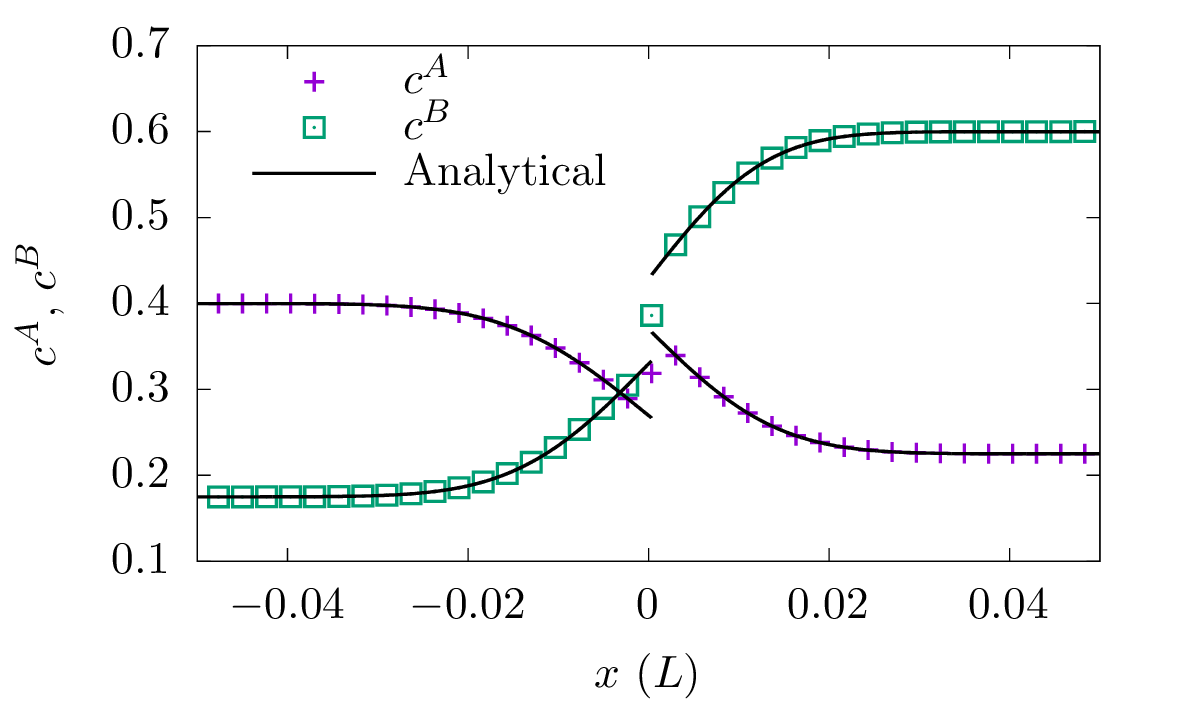}
\par\end{centering}
\caption{\label{fig:LBM_Verif_DiffusionCouple}Benchmark of the LBM code with
an analytical solution for a ternary diffusion couple. (a) Interface displacement
as a function of time, (b) Composition profiles. The symbols are the simulation
results, the lines the analytical predictions. All parameters are listed in 
Table \ref{tab:Parameters_Diffusion-Couple}.}
\end{figure}

To reproduce this problem numerically, the phase-field model is solved
on a thin 2D domain (3000 \texttimes{} 6 LBM lattice nodes). The phase
field is initialized with the equilibrium hyperbolic tangent profile.
The composition fields are initialized as step functions. The parameters used are 
referenced in table \ref{tab:Parameters_Diffusion-Couple} and they produce an analytical
solution with $\xi=-0.269824$. Figure \ref{fig:LBM_Verif_DiffusionCouple}
compares the interface velocity and the concentration fields simulated with the ones 
expected from the analytical solution. Both are in excellent agreement, numerically
confirming the interface condition (\ref{eq:Gibbs-Thomson})--(\ref{eq:MassBalance})
reconstructed by the phase field model. Note that we have chosen the system's half-length
$L$ and the diffusion time $t_{D}=\text{min}_{\alpha}(L^{2}/\dml{M}^{\alpha})$ as
units in these comparisons.

\subsection{Simulations of Ostwald ripening}

Next, the phase-field model is exploited to simulate the Ostwald ripening
of a set of droplets (phase 1) inside a continuous matrix (phase 0). 
As the Allen-Cahn model cannot describe the initial regime of phase separation
from homogeneous mixtures, the initial condition starts from already developed droplets,
as detailed below. In this problem, the characteristic
length scale is the average spacing between droplets. However, this
length is not static and increases as the growth proceeds and the
droplet number decreases. In the numerical simulations, the distance between
droplets is bounded by the domain size. For this reason, we once again
set our numerical units of length and time as the half-length of the
domain and the diffusion time, respectively, $\ell=L$ and 
$t_{D}=\text{min}_{\alpha}(\ell^{2}/\dml{M}^{\alpha})$.

\subsubsection{Initial condition}

As already mentioned above, to study the growth kinetics,
we must start from pre-existing droplets. The initial
condition must be carefully constructed to satisfy the condition for
growth, namely that the global composition inventory lies in the miscibility
gap and that the phase fraction be sufficiently
high to measure a statically relevant average. In addition, before
the growth, there will be a transient regime during which the droplets
reach local equilibrium with the surrounding matrix (analogous to
a diffusion couple). We want the droplets to always tend to grow during
this regime, and not to shrink and risk disappearance.

We therefore choose to parametrize the initialization of the composition
$\boldsymbol{c}(\boldsymbol{x},0)$ and the phase-field $\varphi(\boldsymbol{x},0)$
with three scalar values: the initial global composition of the system
$\boldsymbol{c}_{g}=(c_{g}^{A},c_{g}^{B})$ and the initial phase
fraction $\Phi$ of droplets. The phase-field $\varphi(\boldsymbol{x},0)$
can be easily initialized by creating spherical droplets with randomly
generated positions and radii. For the compositions, we could then
initialize them inside and outside those interfaces using the reference
tie-line\textquoteright s end-points. However, specifying $\Phi$,
$\boldsymbol{c}_{g}$ and the tie-line would overspecify the initialization
because of the lever rule. Besides, if the initial composition of
the matrix is at equilibrium $\boldsymbol{c}_{0}^{eq}$, the droplets
can disappear during the transient regime. It is necessary to design
an initial condition on $c^{A}(\boldsymbol{x},0)$ and $c^{B}(\boldsymbol{x},0)$
such that the droplets grow from the outset of the time evolution.

When the thermodynamic equilibrium is reached, because of the conservation
rule, the global composition $\boldsymbol{c}_{g}$ are related to
the equilibrium compositions and the equilibrium phase fraction $\Phi^{eq}$
by:

\begin{equation}
\boldsymbol{c}_{g}=(1-\Phi^{eq})\boldsymbol{c}_{0}^{eq}+\Phi^{eq}\boldsymbol{c}_{1}^{eq}\label{eq:Global_Compos}
\end{equation}
 That relation can be inverted and yields:

\begin{equation}
\Phi^{eq}=\frac{\left|\boldsymbol{c}_{0}^{eq}-\boldsymbol{c}_{g}\right|}{\left|\boldsymbol{c}_{0}^{eq}-\boldsymbol{c}_{1}^{eq}\right|}\label{eq:phi_glob_eq}
\end{equation}

One way to avoid droplet disappearance is to initialize the
system with a phase fraction of droplets $\Phi$ which is lower than
the equilibrium $\Phi^{eq}$. In that case, we need to initialize
the composition in the matrix and/or in the droplets out of equilibrium
(with an additional degree of freedom due to the ternary case). Here,
the droplet compositions are considered at equilibrium $\boldsymbol{c}_{1}^{eq}$
and we choose to supersaturate the matrix by offsetting its composition
along the tie-line, as

\begin{equation}
\boldsymbol{c}_{0}^{ini}=\boldsymbol{c}_{0}^{eq}-\delta(\boldsymbol{c}_{1}^{eq}-\boldsymbol{c}_{0}^{eq})\label{eq:Init_c0}
\end{equation}
where the coefficient $\delta$ is defined by:

\begin{align}
\delta & =\frac{\Phi^{eq}-\Phi}{1-\Phi}\label{eq:Delta-1}
\end{align}
where $\Phi^{eq}$ is given by Eq. (\ref{eq:phi_glob_eq}). When $\delta>0$
(\emph{i.e.} when $\Phi<\Phi^{eq}$), the matrix composition is offset
inside the miscibility gap. This ensures that the initial transient leading
towards local equilibrium always corresponds to an increase of the global 
phase fraction of droplets $\Phi$. Finally, the initial conditions of compositions write

\begin{equation}
\boldsymbol{c}(\boldsymbol{x},0)=\begin{cases}
(1-\delta)\boldsymbol{c}_{0}^{eq}+\delta\boldsymbol{c}_{1}^{eq} & \text{if }\varphi(\boldsymbol{x})<1/2\\
\boldsymbol{c}_{1}^{eq} & \text{if }\varphi(\boldsymbol{x})\geq1/2
\end{cases}\label{eq:Compos_Init}
\end{equation}
which represents an initial supersaturation in the matrix, whereas
the droplet composition is supposed to be at equilibrium.

\begin{figure}
\begin{centering}
\includegraphics[width=0.90\linewidth]{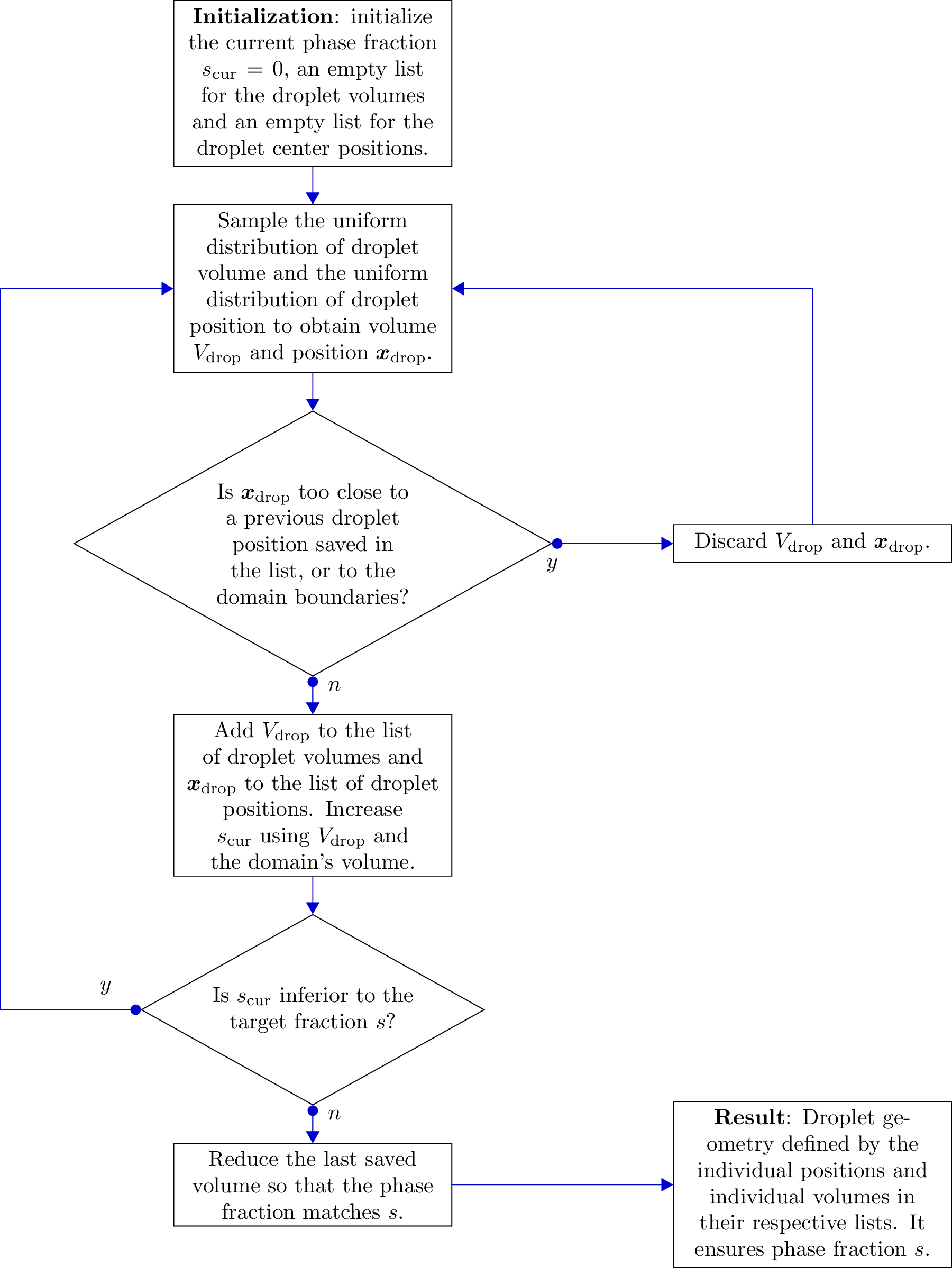}
\par\end{centering}
\caption{\label{fig:initialization_flowchart}Procedure
to initialize the droplet ensemble. Its parameters are the target phase
fraction $s$, the domain's volume, and two uniform distributions for positions
and volumes. A droplet is discarded when the sampled $x_{\text{drop}}$ is
within $R_{\text{max}} + W/2$ to the domain bounds, or within twice that
distance to a previous droplet's center; with $R_{\text{max}}$ the radius
corresponding to the upper bound of the volume distribution.}
\end{figure}

The details of the initialization routine we
used are given next. For a specified target phase fraction $s$,
we initialize the droplet geometry according to the procedure illustrated in
Figure \ref{fig:initialization_flowchart} which results in lists of positions and
volumes for each droplet. Then, to initialize $\varphi(\boldsymbol{x})$ on each
lattice node, search for the closest droplet where we define the hyperbolic
tangent profile using its radius and center. Next, the composition fields
are initialized on each lattice node as Eq. (\ref{eq:Compos_Init}) for a given composition
inventory $\boldsymbol{c}_{g}$, followed by the chemical potential fields
at equilibrium everywhere, $\dml{\mu}^{\alpha}(\boldsymbol{x})=0$. Finally,
for simulations with flow, the flow field is initialized at rest and the pressure
field at zero.

\subsubsection{Geometry measurements}

To measure the droplet count $N$, we use an algorithm for the measure
of the Euler characteristic, to be understood here as the number of
connected regions of $\varphi=1$. The algorithm is adapted from reference
\cite{wiemker_2013} and it only requires a single pass over the
lattice. The average radius $\bigl\langle R\bigr\rangle$ is measured
using the integral estimate

\begin{align}
\int\frac{4}{W}\varphi(1-\varphi)d^{2}\boldsymbol{x} & \approx\sum_{i}2\pi\int_{0}^{+\infty}\frac{4}{W}\varphi_{0}(1-\varphi_{0})rdr\nonumber \\
 & \approx2\pi W\sum_{i}\left[\frac{R_{i}}{W}+\mathcal{O}(e^{-4R_{i}/W})\right]\nonumber \\
 & \approx2\pi N\bigl\langle R\bigr\rangle\label{eq:Rmoy_2D}
\end{align}
where the index $i$ stands for the droplet number. Indeed, the integrand is a function that
has a sharp peak in the interfaces, and the integral approximately gives the total length
of interface present in the system. Assuming the droplets are spherical and sufficiently far
from each other, the integrals can be calculated exactly and a Taylor
expansion gives the desired result, plus an error terms in $\mathcal{O}(e^{-4R_{i}/W})$;
the numerical calculation of this integral is easy to implement 
and very fast (linear in lattice points and no stencil involved).

\subsubsection{Ripening without flow}

\begin{figure}
\begin{centering}
\includegraphics[width=\linewidth]{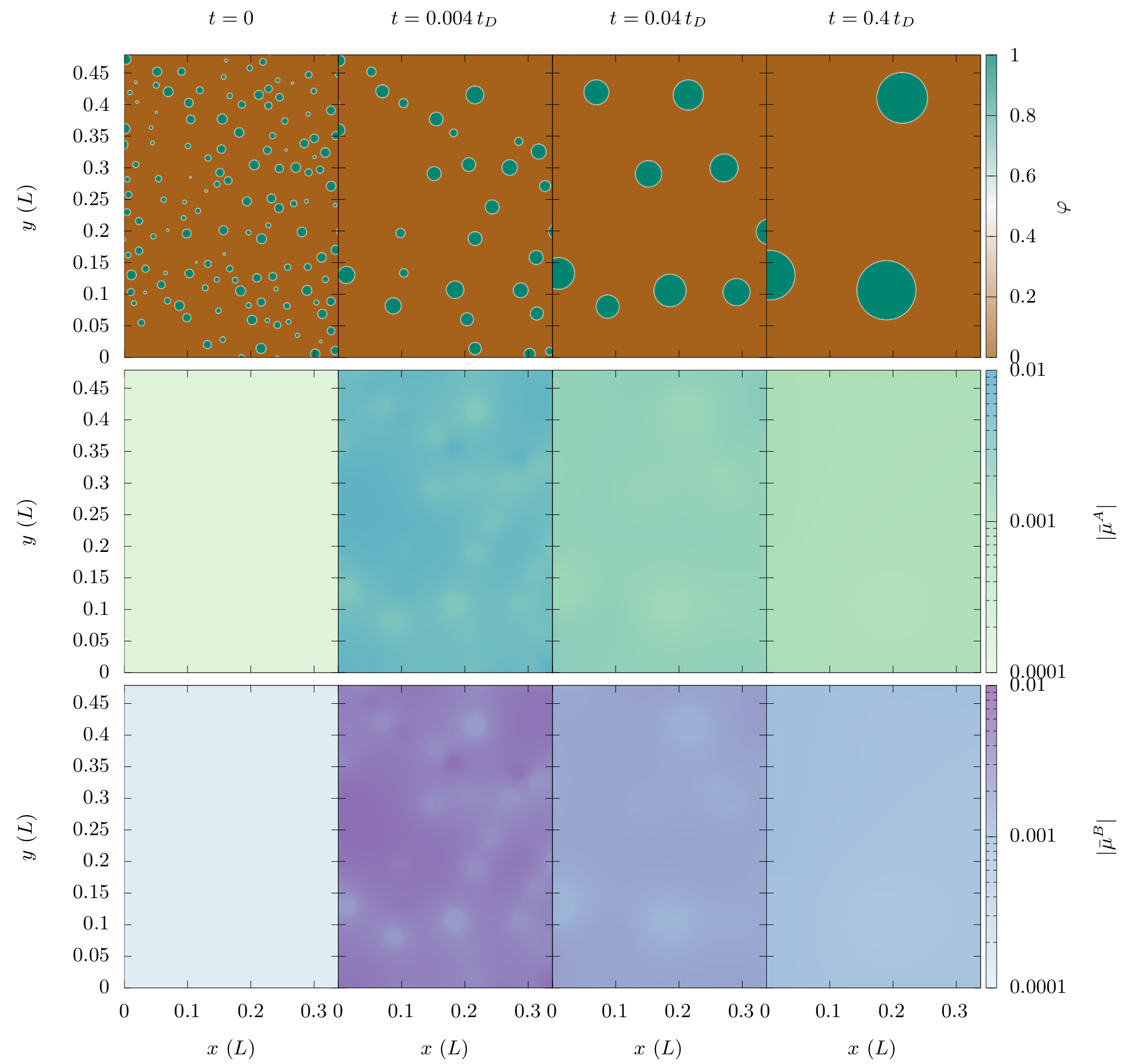}
\par\end{centering}
\caption{\label{fig:Fields_Ostwald-Simulation}Heatmaps of the phase field
and chemical potentials fields in a portion of the domain at various
time steps during the simulation of 2D ripening without flow.}
\end{figure}

We first consider the ripening of a population of two-dimensional droplets without
flow to test our initialization procedure and to reproduce the known growth law
for $\bigl\langle R\bigr\rangle$. The domain ranges between $[-1,1]\times[-1,1]$ and the boundary conditions are periodic. The parameters of the simulation are listed in Table \ref{tab:Parameters_Ostwald-Ripening}.
The initialization produced 2737 droplets. Figure \ref{fig:Fields_Ostwald-Simulation}
shows the phase field and chemical potential fields in a small part of
the domain between $[0,0.3]\times[0,0.45]$ containing only a small fraction of the droplet count. The phase 1 domains follow the expected evolution: the smaller
droplets shrink and vanish while the larger ones grow. The chemical
potentials are initialized at zero (equilibrium). The Gibbs-Thomson
condition creates differences between droplets and their surrounding
matrix and neighbours of different size. As the grains grow and rarify,
the chemical potentials gradually homogeneize. Figure \ref{fig:Radius-Evolution}
presents the evolution of the mean grain radius compared to a power
law with exponent $1/3$. With this, we confirm the adequacy of the
model to reproduce the growth kinetics.

\begin{sidewaystable}
\begin{centering}
\begin{tabular}{lllclllclll}
\cline{1-3} \cline{2-3} \cline{3-3} \cline{5-7} \cline{6-7} \cline{7-7} \cline{9-11} \cline{10-11} \cline{11-11} 
\multicolumn{3}{c}{\textbf{Domain (D2Q9)}} & \hspace{5mm} & \multicolumn{3}{c}{\textbf{Phase-field parameters}} & \hspace{5mm} & \multicolumn{3}{c}{\textbf{Transport parameters}}\tabularnewline
\cline{1-3} \cline{2-3} \cline{3-3} \cline{5-7} \cline{6-7} \cline{7-7} \cline{9-11} \cline{10-11} \cline{11-11} 
Symbol & Value & Dim &  & Symbol & Value & Dim &  & Symbol & Value & Dim\tabularnewline
$[-L_{x},L_{x}]$ & $[-1,1]$ & {[}L{]} &  & $W$ & $3\delta x=1.464\times10^{-3}$ & {[}L{]} &  & $(c_{0}^{A,\eq},c_{0}^{B,\eq})$ & $(0.3,0.3)$ & {[}--{]}\tabularnewline
$[-L_{y},L_{y}]$ & $[-1,1]$ & {[}L{]} &  & $\lambda$ & $155.95$ & {[}--{]} &  & $(c_{1}^{A,\eq},c_{1}^{B,\eq})$ & $(0.4,0.4)$ & {[}--{]}\tabularnewline
$N_{x}$ nodes & $4096$ & {[}--{]} &  & $\dml{M}_{\varphi}$ & $1.2$ & {[}L{]}$^{2}$/{[}T{]} &  & $(\dml{M}_{0}^{AA},\dml{M}_{0}^{BB})$ & $(1,0.8)$ & {[}L{]}$^{2}$/{[}T{]}\tabularnewline
\cline{5-7} \cline{6-7} \cline{7-7} 
$N_{y}$ nodes & $4096$ & {[}--{]} &  & \multicolumn{3}{c}{\textbf{Droplet initialization}} &  & $(\dml{M}_{1}^{AA},\dml{M}_{1}^{BB})$ & $(1,0.8)$ & {[}L{]}$^{2}$/{[}T{]}\tabularnewline
\cline{5-7} \cline{6-7} \cline{7-7} 
$\delta x$ & $4.882\times10^{-4}$ & {[}L{]} &  & $(c_{g}^{A},c_{g}^{B})$ & $(0.31,0.31)$ & {[}--{]} &  &  &  & \tabularnewline
$\delta t$ & $2.5\times10^{-8}$ & {[}T{]} &  & $\Phi$ & $0.08$ & {[}--{]} &  &  &  & \tabularnewline
 &  &  &  & $S_{avg}\pm\Delta S$ & $(1.02\pm0.957)\times10^{-4}$ & {[}L{]}$^{2}$ &  &  &  & \tabularnewline
\cline{1-3} \cline{2-3} \cline{3-3} \cline{5-7} \cline{6-7} \cline{7-7} \cline{9-11} \cline{10-11} \cline{11-11} 
\end{tabular}
\par\end{centering}
\caption{\label{tab:Parameters_Ostwald-Ripening}Parameters for the simulation
of 2D Ostwald ripening. The initial volumes of the grains are uniformly
sampled in the range $[S_{avg}-\Delta S,S_{avg}+\Delta S]$. The values of collision rates corresponding respectively to $\dml{M}_{\varphi}$, $\dml{M}^{AA}$ and $\dml{M}^{BB}$ are: $\tau_{g}=0.8776$, $\tau_{h}^{A}=0.8146$ and $\tau_{h}^{B}=0.7517$.}
\end{sidewaystable}

\begin{center}
\begin{figure}
\begin{centering}
\includegraphics[height=5cm]{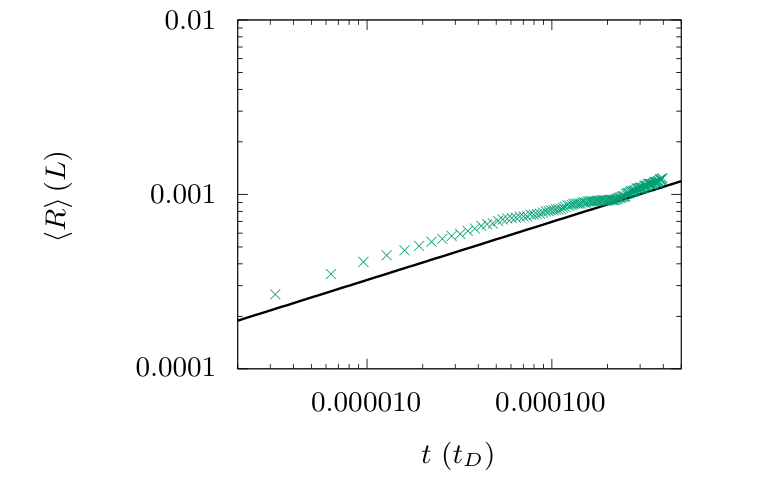}
\par\end{centering}
\caption{\label{fig:Radius-Evolution}Mean droplet radius versus time in
logarithmic scales during the Ostwald ripening simulation without flow. The green crosses
are the radii measured from the simulation. The black line is a reference
line for the relation $\bigl\langle R\bigr\rangle\propto t^{1/3}$.}
\end{figure}
\par\end{center}

\subsubsection{Ripening and sedimentation}

Next, we look at a three-dimensional ripening process including the effect of buoyancy.
Table \ref{tab:Parameters_RipeningFlow} lists the
parameters. The gravity acts along the $x$-axis and the bounce-back
LBM algorithm is used on the top and bottow walls, enforcing $\boldsymbol{u}=\boldsymbol{0}$
and $\boldsymbol{\nabla}\varphi=\boldsymbol{\nabla}c=\boldsymbol{\nabla}\dml{\mu}=\boldsymbol{0}$.
The lateral boundaries are periodic.

\begin{sidewaystable}
\begin{centering}
\begin{tabular}{lllclllclll}
\cline{1-3} \cline{2-3} \cline{3-3} \cline{5-7} \cline{6-7} \cline{7-7} \cline{9-11} \cline{10-11} \cline{11-11} 
\multicolumn{3}{c}{\textbf{Domain (D3Q19)}} & \hspace{5mm} & \multicolumn{3}{c}{\textbf{Phase-field parameters}} & \hspace{5mm} & \multicolumn{3}{c}{\textbf{Droplet initialization}}\tabularnewline
\cline{1-3} \cline{2-3} \cline{3-3} \cline{5-7} \cline{6-7} \cline{7-7} \cline{9-11} \cline{10-11} \cline{11-11} 
Symbol & Value & Dim &  & Symbol & Value & Dim &  & Symbol & Value & Dim\tabularnewline
$[-L_{x},L_{x}]$ & $[-16,16]$ & {[}L{]} &  & $W$ & $4\delta x=6.25\times10^{-2}$ & {[}L{]} &  & $(c_{g}^{A},c_{g}^{B})$ & $(0.32,0.32)$ & {[}--{]}\tabularnewline
$[-L_{y},L_{y}]$ & $[-4,4]$ & {[}L{]} &  & $\lambda$ & $155.95$ & {[}--{]} &  & $\Phi$ & $0.08$ & {[}--{]}\tabularnewline
$[-L_{z},L_{z}]$ & $[-4,4]$ & {[}L{]} &  & $\dml{M}_{\varphi}$ & $1.2$ & {[}L{]}$^{2}$/{[}T{]} &  & $V_{avg}\pm\Delta V$ & $0.08\pm0.072$ & {[}L{]}$^{3}$\tabularnewline
\cline{5-7} \cline{6-7} \cline{7-7} \cline{9-11} \cline{10-11} \cline{11-11} 
$N_{x}$ nodes & $2048$ & {[}--{]} &  & \multicolumn{3}{c}{\textbf{Flow parameters}} &  & \multicolumn{3}{c}{\textbf{Transport parameters}}\tabularnewline
\cline{5-7} \cline{6-7} \cline{7-7} \cline{9-11} \cline{10-11} \cline{11-11} 
$N_{y}$ nodes & $512$ & {[}--{]} &  & $\rho_0=\Delta\rho$ & $1$ & {[}M{]}/{[}L{]}$^{3}$ &  & $(c_{0}^{A,\eq},c_{0}^{B,\eq})$ & $(0.3,0.3)$ & {[}--{]}\tabularnewline
$N_{z}$ nodes & $512$ & {[}--{]} &  & $\nu$ & $1$ & {[}L{]}$^{2}$/{[}T{]} &  & $(c_{1}^{A,\eq},c_{1}^{B,\eq})$ & $(0.4,0.4)$ & {[}--{]}\tabularnewline
$\delta x$ & $1.5625\times10^{-2}$ & {[}L{]} &  & $\sigma$ & $10^{-3}$ & {[}MLT$^{-2}${]}/{[}L{]} &  & $(\dml{M}_{0}^{AA},\dml{M}_{0}^{BB})$ & $(1,0.8)$ & {[}L{]}$^{2}$/{[}T{]}\tabularnewline
$\delta t$ & $4.2\times10^{-5}$ & {[}T{]} &  & $\Delta\rho g_{x}$ & $1$ & {[}M{]}/{[}LT{]}$^{2}$ &  & $(\dml{M}_{1}^{AA},\dml{M}_{1}^{BB})$ & $(1,0.8)$ & {[}L{]}$^{2}$/{[}T{]}\tabularnewline
\cline{1-3} \cline{2-3} \cline{3-3} \cline{5-7} \cline{6-7} \cline{7-7} \cline{9-11} \cline{10-11} \cline{11-11} 
\end{tabular}
\par\end{centering}
\caption{\label{tab:Parameters_RipeningFlow}Parameters for the simulation
of 3D Ostwald ripening with a gravity-driven flow. The initial grains
volumes are uniformly sampled in the range $[V_{avg}-\Delta V,V_{avg}+\Delta V]$. The values of collision rates corresponding respectively to $\nu$, $\dml{M}_{\varphi}$, $\dml{M}^{AA}$ and $\dml{M}^{BB}$ are: $\tau_{f}=0.6111$, $\tau_{g}=0.6333$, $\tau_{h}^{A}=0.6111$ and $\tau_{h}^{B}=0.5888$.}
\end{sidewaystable}

\begin{figure}
\begin{centering}
\subfloat[\label{fig:Ostwald_0}$\overline{t}_{0}=0$]{
\centering{}\includegraphics[height=6.5cm]{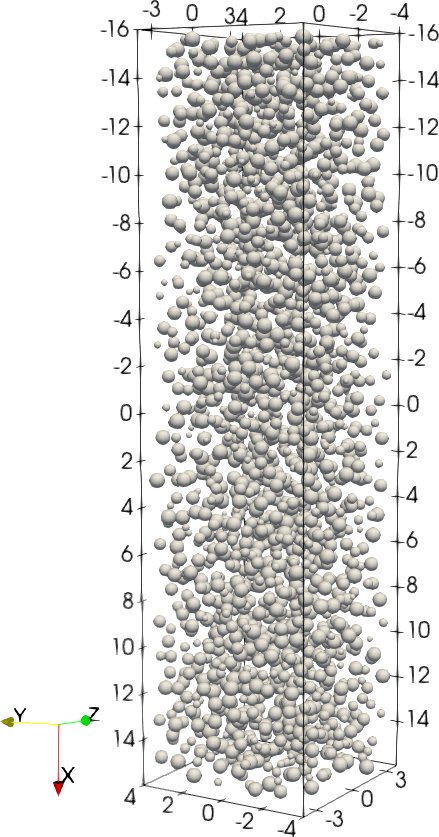}}\hspace{3mm}\subfloat[\label{fig:Ostwald_1}$\overline{t}_{1}=0.966$]{
\centering{}\includegraphics[height=6.5cm]{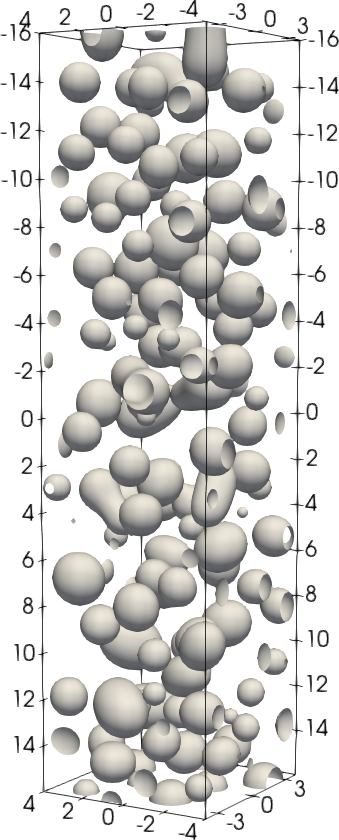}}\hspace{3mm}\subfloat[\label{fig:Ostwald_2}$\overline{t}_{2}=2.814$]{
\centering{}\includegraphics[height=6.5cm]{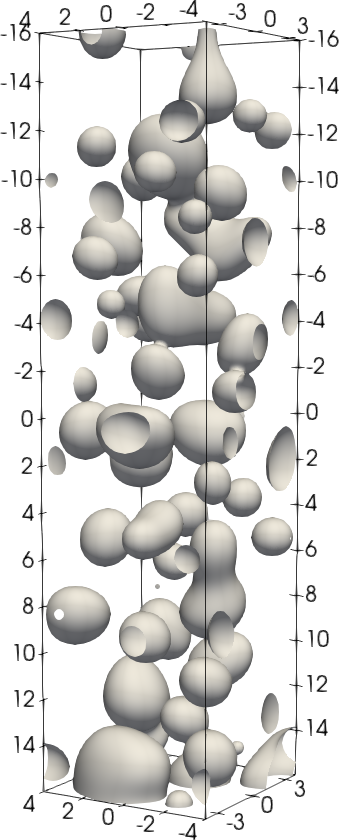}}\hspace{3mm}\subfloat[\label{fig:Ostwald_3}$\overline{t}_{3}=4.2$]{
\centering{}\includegraphics[height=6.5cm]{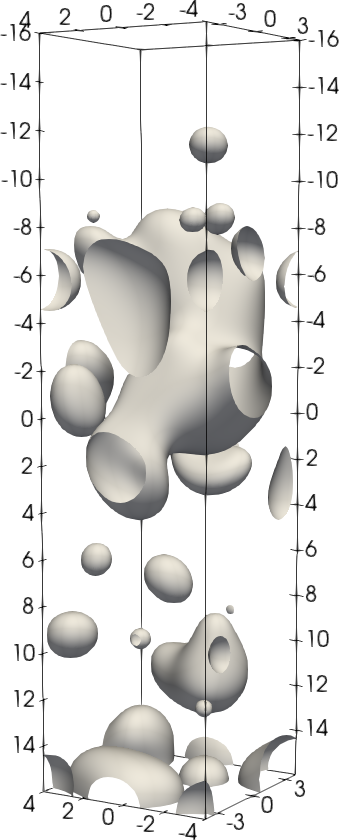}}\hspace{3mm}\subfloat[\label{fig:Ostwald_4}$\overline{t}_{4}=6.216$]{
\centering{}\includegraphics[height=6.5cm]{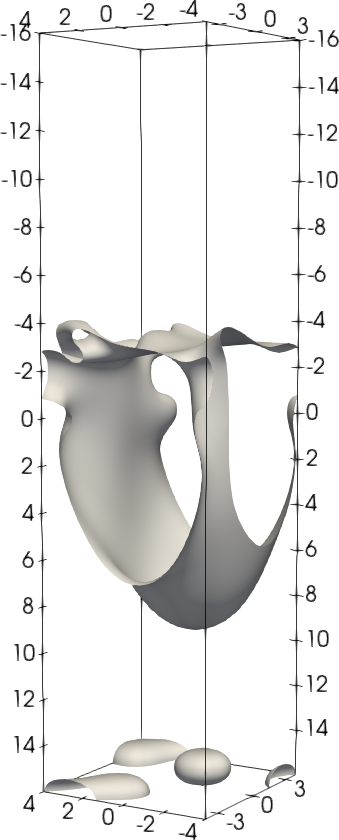}}
\par\end{centering}
\caption{\label{fig:3D-simulation_OstwaldFlow}Three-dimensional simulation of droplet 
ripening with sedimentation caused by a constant gravity oriented downwards.
Snapshots of the interface ($\varphi=0.5$) at several times $\overline{t}_{i}=t_{i}/t_{D}$
where $t_{D}$ is the diffusion time $t_{D}=\ell^{2}/\dml{M}^{A}$
with $\ell=4$ and $\dml{M}^{A}=1$. The simulation outputs $t_{i}$
are ($\times10^{5}\delta t$): $t_{1}=3.68$, $t_{2}=10.72$, $t_{3}=16$
and $t_{4}=23.68$. The simulation took 24h on 64 GPU A100.}
\end{figure}

\begin{figure}
\begin{centering}
\includegraphics[height=9cm]{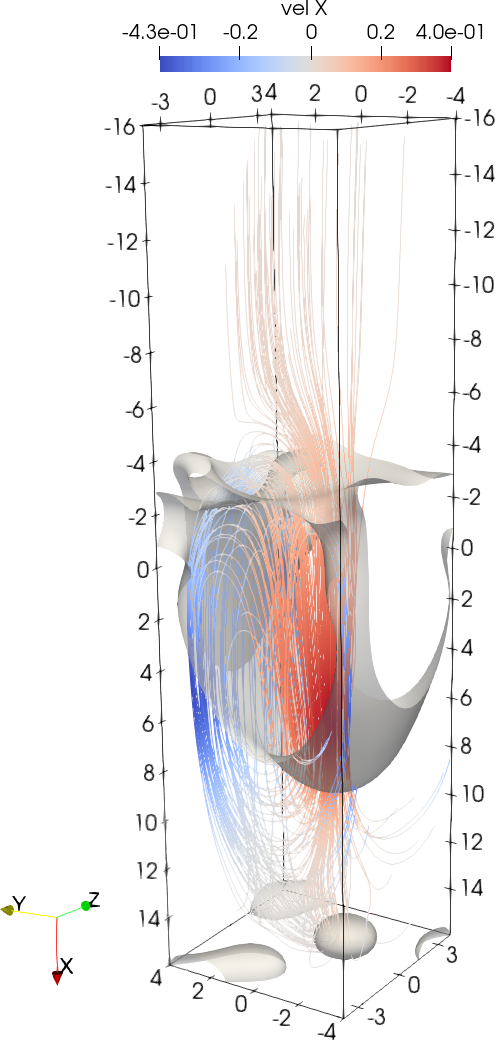}
\par\end{centering}
\caption{\label{fig:Streamlines}Streamlines at the final time
of droplet ripening with sedimentation. The streamlines are colored
by the velocity component along the $x$-axis, showing the recirculation
around the large falling drop.}

\end{figure}

Figure \ref{fig:3D-simulation_OstwaldFlow} shows successive snapshot pictures of
the droplet geometry. The initial condition (Figure \ref{fig:Ostwald_0})
is composed of 2035 droplets. At the beginning, diffusive Ostwald ripening
takes place, and after some time (Figure \ref{fig:Ostwald_1}), the smallest droplets 
have vanished and the remaining ones start to sediment. The droplets then 
accelerate (Figure \ref{fig:Ostwald_2}), and some have already reached the bottom
wall. A droplet can be seen hanging from the top wall due to capillarity.
Many droplets lose their spherical shape due to coalescence events. Much
later (Figure \ref{fig:Ostwald_3}), the top of the domain contains almost
no droplets anymore due to the combined effect of ripening, sedimentation
and coalescence. A large drop of the dense phase forms. This huge mass
continues to sediment, but the surrounding smaller droplets rise due to a recirculation
of the flow field. At the final time of the simulation (Figure \ref{fig:Ostwald_4}),
only a few droplets are left at the bottom. They are still subject
to evaporation due to the very large drop that formed in the center.
The flow recirculation is clearly seen at the final time step on Figure
\ref{fig:Streamlines}. At earlier times, small droplets may also
be seen rising around the large domain of phase 1 before they quickly
evaporate.

\begin{figure}
\begin{centering}
\includegraphics[height=4.5cm]{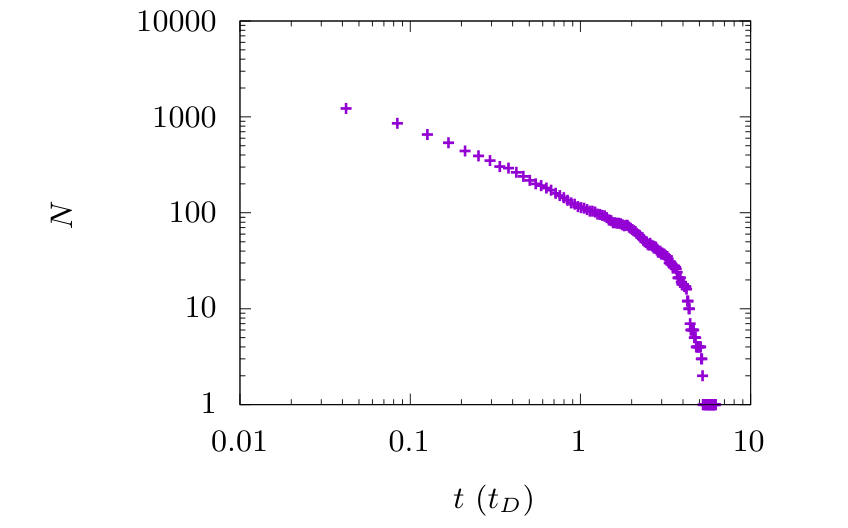}
\hspace{5mm}
\includegraphics[height=4.5cm]{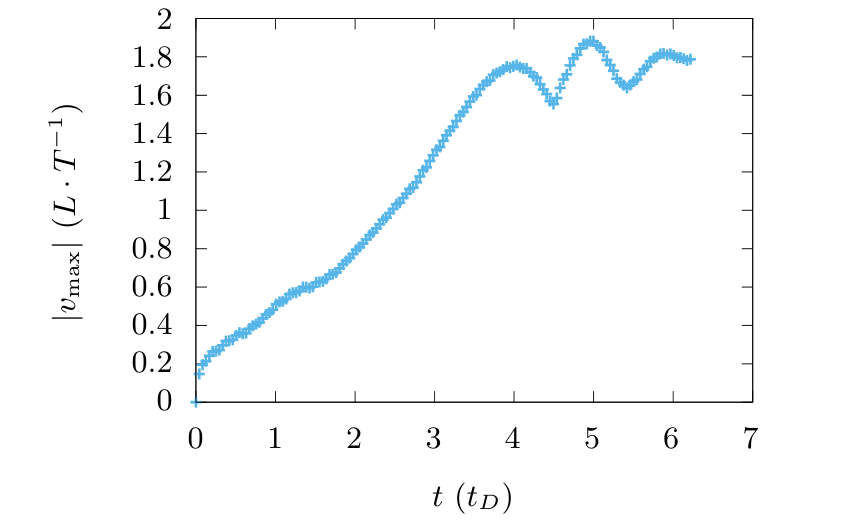}
\par\end{centering}
\caption{\label{fig:Time-evolution}Droplet count (a) and maximum fluid velocity (b) as a function of time during the
simulation of Figure~\ref{fig:3D-simulation_OstwaldFlow}.}
\end{figure}

Figure~\ref{fig:Time-evolution}(a) tracks the number of droplets in the simulation as a
function of time. Around $t=4t_{D}$, the very large drop 
forms in the center and its size becomes comparable to the system size.
Hence, this single drop totally dominates the droplet distribution: the remaining droplets
are few in number, and are very small in comparison (as seen in Figure \ref{fig:3D-simulation_OstwaldFlow}).
Because of the small system size and since all small droplets
evaporate in favor of the largest drop, the growth dynamics rapidly stops to obey a power law.
The development of the sedimentation dynamics is illustrated in Figure~\Ref{fig:Time-evolution}(b),
which displays the maximum fluid velocity in the domain as a function of time. In the
beginning, this maximum velocity increases monotonically: the average droplet size 
increases with time, which leads to higher sedimentation velocities. After the formation
of the large drop at $t=4t_{D}$, the velocity saturates and oscillates. In this regime,
the finite system size limits the sedimantation velocity of the central
drop, and the highest velocities occur in the upward recirculation.

\section{Conclusion and outlook}
We have constructed and tested a model for the simulation of late-stage coarsening in
phase-separating liquids, taking into account buoyancy effects which lead to droplet
sedimentation. Two-phase coexistence and the motion of interfaces are described by
a grand-canonical phase-field model, which can be adapted to arbitrary thermodynamic
properties as described by a free-energy model. The coupled model is simulated
numerically by the lattice-Boltzmann method, which is used to integrate in time
all equations of the model. This makes the simulation code monolithic despite its
multiphysics nature. Therefore, it is particularly well adapted for high-performance
simulations on GPU architectures.

We have demonstrated the capabilities of our approach
by several test cases. In Section \ref{Simuls} we started by verifying that this model reconstructs the correct coupling between interface kinetics and thermodynamics of a three-components system when compared with the equivalent sharp-interface problem. Next, simulations of Ostwald ripening have been carried out in two dimensions in a large system, and we have verified that the mean droplet radius follows the expected power law with time. Finally, one large-scale three-dimensional simulation of simultaneous coarsening and droplet sedimentation has been performed. It has demonstrated that our model can be
used to study the interplay between coarsening and fluid flow created by droplet sedimentation.

With respect to a multi-component Cahn-Hilliard model, our approach has at least
two advantages: as already stated, it can be adapted to simulate any desired substance
for which free-energy data are available, without restrictions on the choice of
the surface tension. Furthermore, the equations of the phase-field model are second
order in space, rather than the fourth order Cahn-Hilliard equation, which makes its
numerical integration more efficient. The price to pay is that the initial stages of
phase separation cannot be described by this model, which makes it necessary to construct
an initial state {\em ad hoc} by making hypotheses on the initial droplet size distribution.
However, since the memory of the initial state is rapidly lost during the coarsening of
a disordered droplet distribution, this should not be a severe problem.

We have written down a model for a ternary mixture here, but there is no difficulty
in generalizing the approach to mixtures with more than three components, as long
as the necessary thermodynamic and kinetic data (free energies and mobilities)
are available. An interesting question that could be studied in the future is the
interplay between coarsening and sedimentation in the macroscopic redistribution
of components: since the sedimentation of the droplets leads to an accumulation
of the denser phase at the bottom of the system, and the sedimentation velocity
is linked to the droplet size, there is a non-trivial coupling between the two
phenomena.

\section*{Acknowledgements}

This work was granted access to the HPC resources of IDRIS (super-computer Jean-Zay, partition V100) and CCRT (Topaze, partition A100). Alain Cartalade wishes to thank the SIVIT project involving Orano and EDF for the financial support.


\begin{thebibliography}{10}

\bibitem{gunton1983phase}
JD~Gunton, M~San~Miguel, and PS Sahni.
\newblock The dynamics of first-order phase transitions.
\newblock In C.~Domb and J.~L.~Lebowitz, editors, {\em Phase Transitions and Critical Phenomena, Vol. 8}, pages 267--466, New York, 1983. Academic Press.

\bibitem{siggia_1979}
Eric~D. Siggia.
\newblock Late stages of spinodal decomposition in binary mixtures.
\newblock {\em Phys. Rev. A}, 20:595--605, Aug 1979.

\bibitem{Bray94}
A.J. Bray.
\newblock Theory of phase-ordering kinetics.
\newblock {\em Advances in Physics}, 43(3):357--459, 1994.

\bibitem{Bray02}
A.~J. Bray.
\newblock Theory of phase-ordering kinetics.
\newblock {\em Advances in Physics}, 51(2):481--587, 2002.

\bibitem{gin2017radionuclides}
Stephane Gin, Patrick Jollivet, Magaly Tribet, Sylvain Peuget, and Sophie
  Schuller.
\newblock Radionuclides containment in nuclear glasses: an overview.
\newblock {\em Radiochimica Acta}, 105(11):927--959, 2017.

\bibitem{Schuller_etal_JACS2011}
Sophie Schuller, Olivier Pinet, and Bruno Penelon.
\newblock Liquid–liquid phase separation process in borosilicate liquids
  enriched in molybdenum and phosphorus oxides.
\newblock {\em Journal of the American Ceramic Society}, 94(2):447--454, 2011.

\bibitem{Pinet_etal_JNM2019}
O.~Pinet, J.-F. Hollebecque, I.~Hugon, V.~Debono, L.~Campayo, C.~Vallat, and
  V.~Lemaitre.
\newblock Glass ceramic for the vitrification of high level waste with a high
  molybdenum content.
\newblock {\em Journal of Nuclear Materials}, 519:121--127, 2019.

\bibitem{Cahn58}
J.~W. Cahn and J.~E. Hilliard.
\newblock Free energy of a non-uniform system. 1. interfacial free energy.
\newblock {\em J. Chem. Phys.}, 28:258--267, 1958.

\bibitem{brackbill_kothe_zemach_1992}
J.U Brackbill, D.B Kothe, and C~Zemach.
\newblock A continuum method for modeling surface tension.
\newblock {\em Journal of Computational Physics}, 100(2):335--354, 1992.

\bibitem{jacqmin_1999}
David Jacqmin.
\newblock Calculation of two-phase navier–stokes flows using phase-field
  modeling.
\newblock {\em Journal of Computational Physics}, 155(1):96--127, 1999.

\bibitem{henry_tegze_2018}
Herv\'e Henry and Gy\"orgy Tegze.
\newblock Self-similarity and coarsening rate of a convecting bicontinuous
  phase separating mixture: Effect of the viscosity contrast.
\newblock {\em Phys. Rev. Fluids}, 3:074306, Jul 2018.

\bibitem{henry_tegze_2019}
Herv\'e Henry and Gy\"orgy Tegze.
\newblock Kinetics of coarsening have dramatic effects on the microstructure:
  Self-similarity breakdown induced by viscosity contrast.
\newblock {\em Phys. Rev. E}, 100:013116, Jul 2019.

\bibitem{semprebon_ciro_kruger_2016}
Ciro Semprebon, Timm Kr\"uger, and Halim Kusumaatmaja.
\newblock Ternary free-energy lattice boltzmann model with tunable surface
  tensions and contact angles.
\newblock {\em Phys. Rev. E}, 93:033305, Mar 2016.

\bibitem{rasolofomanana_et_al_2022}
M.A. Rasolofomanana, C.~Cardon, M.~Plapp, T.~Philippe, H.~Henry, and R.~{Le
  Tellier}.
\newblock Diffuse-interface modelling of multicomponent diffusion and phase
  separation in the u-o-zr ternary system.
\newblock {\em Computational Materials Science}, 214:111650, 2022.

\bibitem{ProvatasElder}
N.~Provatas and K.~Elder.
\newblock {\em Phase-field methods in materials science and engineering}.
\newblock Wiley-VCH, Weinheim, 2010.

\bibitem{Steinbach09}
I.~Steinbach.
\newblock {Phase-field models in materials science}.
\newblock {\em Model. Simul. Mater. Sci. Eng.}, {17}({7}):073001, {2009}.

\bibitem{PlappHandbook}
M.~Plapp.
\newblock Phase-field models.
\newblock In T.~Nishinaga, editor, {\em The Handbook of Crystal Growth, 2nd
  edition, Vol. 1B}, pages 631--668, Amsterdam, 2015. Elsevier.

\bibitem{karma_rappel_1998}
Alain Karma and Wouter-Jan Rappel.
\newblock Quantitative phase-field modeling of dendritic growth in two and
  three dimensions.
\newblock {\em Phys. Rev. E}, 57:4323--4349, Apr 1998.

\bibitem{almgren_1999}
Robert~F. Almgren.
\newblock Second-order phase field asymptotics for unequal conductivities.
\newblock {\em SIAM Journal on Applied Mathematics}, 59(6):2086--2107, 1999.

\bibitem{echebarria_et_al_2004}
Blas Echebarria, Roger Folch, Alain Karma, and Mathis Plapp.
\newblock Quantitative phase-field model of alloy solidification.
\newblock {\em Phys. Rev. E}, 70:061604, Dec 2004.

\bibitem{badillo_2012}
Arnoldo Badillo.
\newblock Quantitative phase-field modeling for boiling phenomena.
\newblock {\em Phys. Rev. E}, 86:041603, Oct 2012.

\bibitem{kim_kim_suzuki_1999}
Seong~Gyoon Kim, Won~Tae Kim, and Toshio Suzuki.
\newblock Phase-field model for binary alloys.
\newblock {\em Phys. Rev. E}, 60:7186--7197, Dec 1999.

\bibitem{plapp_2011_2}
Mathis Plapp.
\newblock Unified derivation of phase-field models for alloy solidification
  from a grand-potential functional.
\newblock {\em Phys. Rev. E}, 84:031601, Sep 2011.

\bibitem{Choudhury12}
A.~Choudhury and B.~Nestler.
\newblock Grand-potential formulation for multicomponent phase transformations
  combined with thin-interface asymptotics of the double-obstacle potential.
\newblock {\em Phys. Rev. E}, 85:021602, 2012.

\bibitem{Plapp16}
M.~Plapp.
\newblock Phase-field modelling of solidification microstructures.
\newblock {\em Journal of the Indian Institute of Science}, 96(3):179--198,
  2016.

\bibitem{the_lattice_boltzmann_method}
Timm Kr{\"u}ger, Halim Kusumaatmaja, Alexandr Kuzmin, Orest Shardt, Goncalo
  Silva, and Erlend~Magnus Viggen.
\newblock {\em The Lattice Boltzmann Method: Principles and Practice}.
\newblock Springer International Publishing, Cham, 2017.

\bibitem{calphad}
Larry Kaufman and Henry~L. Bernstein.
\newblock {\em Computer calculation of phase diagrams with special reference to
  refractory metals}.
\newblock New York : Academic Press, 1970.

\bibitem{cartalade2016}
Alain Cartalade, Amina Younsi, and Mathis Plapp.
\newblock Lattice boltzmann simulations of 3d crystal growth: Numerical schemes
  for a phase-field model with anti-trapping current.
\newblock {\em Computers \& Mathematics with Applications}, 71(9):1784--1798,
  2016.

\bibitem{verdier_kestener_cartalade_2020}
Werner Verdier, Pierre Kestener, and Alain Cartalade.
\newblock Performance portability of lattice boltzmann methods for two-phase
  flows with phase change.
\newblock {\em Computer Methods in Applied Mechanics and Engineering},
  370:113266, 2020.

\bibitem{LangerBegRohu}
J.~S. Langer.
\newblock An introduction to the kinetics of first-order phase transitions.
\newblock In C.~Godr\`eche, editor, {\em Solids far from equilibrium}, Edition
  Al\'ea Saclay, pages 297--363, Cambridge, UK, 1991. Cambridge University
  Press.

\bibitem{bayle_2020}
R.~Bayle, O.~Cueto, S.~Blonkowski, T.~Philippe, H.~Henry, and M.~Plapp.
\newblock Phase-field modeling of the non-congruent crystallization of a
  ternary {Ge}–{Sb}–{Te} alloy for phase-change memory applications.
\newblock {\em Journal of Applied Physics}, 128(18):185101, November 2020.

\bibitem{bayle_2020_phd}
Raphael Bayle.
\newblock {\em Simulation des m\'ecanismes de changement de phase dans des
  m\'emoires PCM avec la m\'ethode multi-champ de phase}.
\newblock PhD thesis, 2020.
\newblock Th\`ese de doctorat dirig\'ee par Plapp, Mathis, Institut
  polytechnique de Paris 2020.

\bibitem{folch_et_al_1999}
R.~Folch, J.~Casademunt, A.~Hern\'andez-Machado, and L.~Ram\'{\i}rez-Piscina.
\newblock Phase-field model for hele-shaw flows with arbitrary viscosity
  contrast. ii. numerical study.
\newblock {\em Phys. Rev. E}, 60:1734--1740, Aug 1999.

\bibitem{sun_beckermann_2007}
Y.~Sun and C.~Beckermann.
\newblock Sharp interface tracking using the phase-field equation.
\newblock {\em Journal of Computational Physics}, 220(2):626--653, jan 2007.

\bibitem{He-Shan-Doolen_PRE-Rapid1998}
Xiaoyi He, Xiaowen Shan, and Gary~D. Doolen.
\newblock Discrete boltzmann equation model for nonideal gases.
\newblock {\em Phys. Rev. E}, 57:R13--R16, Jan 1998.

\bibitem{He-Luo_Incompressible_JSP1997}
X.~He and L.-S. Luo.
\newblock Lattice boltzmann model for the incompressible navier-stokes
  equation.
\newblock {\em Journal of Statistical Physics}, 88(3/4):pp. 927--944, 1997.

\bibitem{zu_he_2013}
Y.~Q. Zu and S.~He.
\newblock Phase-field-based lattice boltzmann model for incompressible binary
  fluid systems with density and viscosity contrasts.
\newblock {\em Phys. Rev. E}, 87:043301, Apr 2013.

\bibitem{Fakhari_etal_PRE2017}
Abbas Fakhari, Travis Mitchell, Christopher Leonardi, and Diogo Bolster.
\newblock Improved locality of the phase-field lattice-boltzmann model for
  immiscible fluids at high density ratios.
\newblock {\em Phys. Rev. E}, 96:053301, Nov 2017.

\bibitem{fakhari_rahimian_2010}
Abbas Fakhari and Mohammad~H. Rahimian.
\newblock Phase-field modeling by the method of lattice boltzmann equations.
\newblock {\em Phys. Rev. E}, 81:036707, Mar 2010.

\bibitem{LBMsaclaycode}
\texttt{LBM\_Saclay} code.
\newblock \texttt{https://gitlab.maisondelasimulation.fr/users/sign\_in}, 2018.

\bibitem{kokkos}
H.~Carter Edwards, Christian~R. Trott, and Daniel Sunderland.
\newblock Kokkos: Enabling manycore performance portability through polymorphic
  memory access patterns.
\newblock {\em Journal of Parallel and Distributed Computing}, 74(12):3202 --
  3216, 2014.
\newblock Domain-Specific Languages and High-Level Frameworks for
  High-Performance Computing.

\bibitem{heulens_blanpain_moelans_2011}
J.~Heulens, B.~Blanpain, and N.~Moelans.
\newblock Phase-field analysis of a ternary two-phase diffusion couple with
  multiple analytical solutions.
\newblock {\em Acta Materialia}, 59(10):3946 -- 3954, 2011.

\bibitem{lahiri_abinandanan_choudhury_2017}
Arka Lahiri, T.~A. Abinandanan, and Abhik Choudhury.
\newblock Theoretical and numerical study of growth in multi-component alloys.
\newblock {\em Metallurgical and Materials Transactions A}, 48:4463--4476,
  October 2017.

\bibitem{maugis_et_al_1997}
P.~Maugis, W.D. Hopfe, J.E. Morral, and J.S. Kirkaldy.
\newblock Multiple interface velocity solutions for ternary biphase infinite
  diffusion couples.
\newblock {\em Acta Materialia}, 45(5):1941 -- 1954, 1997.

\bibitem{wiemker_2013}
Rafael Wiemker.
\newblock {Total Euler Characteristic as a Noise Measure to aid Transfer
  Function Design}.
\newblock In Mario Hlawitschka and Tino Weinkauf, editors, {\em EuroVis - Short
  Papers}. The Eurographics Association, 2013.

\end{thebibliography}

\end{document}